\documentclass[preprint,1p, number]{elsarticle}

\usepackage{framed,multirow}

\usepackage{amsmath}
\usepackage{amsthm}
\usepackage{amssymb}
\usepackage{latexsym}

\usepackage{url}
\usepackage{xcolor}
\definecolor{newcolor}{rgb}{.8,.349,.1}

\usepackage{float}
\floatstyle{plaintop}
\restylefloat{table}
\usepackage{subcaption}
\usepackage{graphicx}

\usepackage{tabularx}
\usepackage{multirow}
\usepackage{booktabs}
\usepackage{lipsum}

\newtheorem{proposition}{Proposition}[section]
\newtheorem{definition}{Definition}[section]

\newcommand{\sgn}[1]{\text{sgn}(#1)}

\usepackage{tikz, pgfplots}
\pgfplotsset{compat=1.18}

\newcommand{\logLogSlopeTriangle}[5]
{

    \pgfplotsextra
    {
        \pgfkeysgetvalue{/pgfplots/xmin}{\xmin}
        \pgfkeysgetvalue{/pgfplots/xmax}{\xmax}
        \pgfkeysgetvalue{/pgfplots/ymin}{\ymin}
        \pgfkeysgetvalue{/pgfplots/ymax}{\ymax}

        \pgfmathsetmacro{\xArel}{#1}
        \pgfmathsetmacro{\yArel}{#3}
        \pgfmathsetmacro{\xBrel}{#1-#2}
        \pgfmathsetmacro{\yBrel}{\yArel}
        \pgfmathsetmacro{\xCrel}{\xArel}

        \pgfmathsetmacro{\lnxB}{\xmin*(1-(#1-#2))+\xmax*(#1-#2)} 
        \pgfmathsetmacro{\lnxA}{\xmin*(1-#1)+\xmax*#1} 
        \pgfmathsetmacro{\lnyA}{\ymin*(1-#3)+\ymax*#3} 
        \pgfmathsetmacro{\lnyC}{\lnyA+#4*(\lnxA-\lnxB)}
        \pgfmathsetmacro{\yCrel}{\lnyC-\ymin)/(\ymax-\ymin)} 

        \coordinate (A) at (rel axis cs:\xArel,\yArel);
        \coordinate (B) at (rel axis cs:\xBrel,\yBrel);
        \coordinate (C) at (rel axis cs:\xCrel,\yCrel);

        \draw[#5]   (A)-- node[pos=0.5,anchor=north] {1}
                    (B)-- 
                    (C)-- node[pos=0.5,anchor=west] {#4}
                    cycle;
    }
}

\begin{document}

\begin{frontmatter}

\title{Enhanced Runge-Kutta Discontinuous Galerkin Method for Ultrasound Propagation in Transit-Time Flow Meters}%

\author[1]{Matteo Calafà\corref{cor1}\fnmark[cor2]}
\ead{maca@mpe.au.dk}
\cortext[cor1]{Corresponding author}
\fntext[cor2]{Formerly at: Institute of Mathematics, École Polytechnique Fédérale de Lausanne (EPFL), 1015 Lausanne, Switzerland}
\author[2]{Martino Reclari}
\ead{marr@kamstrup.com}

\address[1] {Institut for Mekanik \& Produktion, Aarhus University, 8000 Aarhus, Denmark}
\address[2]{Kamstrup A/S, 8660 Skanderborg, Denmark}

\begin{abstract}
We illustrate a time and memory efficient application of Runge-Kutta discontinuous Galerkin (RKDG) methods for the simulation of the ultrasounds advection in moving fluids. In particular, this study addresses to the analysis of transit-time ultrasonic meters which rely on the propagation of acoustic waves to measure fluids flow rate. Accurate and efficient simulations of the physics related to the transport of ultrasounds are therefore crucial for studying and enhancing these devices. Starting from the description of the linearized Euler equations (LEE) model and presenting the general theory of explicit-time DG methods for hyperbolic systems, we then motivate the use of a spectral basis and introduce a novel high-accuracy method for the imposition of absorbing and resistive walls which analyses the incident wave direction across the boundary surface. The proposed implementation is both accurate and efficient, making it suitable for industrial applications of acoustic wave propagation.
\end{abstract}


\end{frontmatter}


 \section{Introduction} \label{introduction} 
High-accuracy simulations have increasingly become essential in many science fields such as fluid dynamics \cite{bassi1997,wang2013high}, electromagnetism \cite{hesthaven2003high}, electrophysiology \cite{belhamadia2009towards,coudiere2017very} and acoustics \cite{lockard1995high}. The flow metering industry has followed this trend and several investigations have been performed aiming at improving and optimizing measurement performances \cite{mubarok2020comparative, saboohi2015developing,rincon2022turbulent}. In particular, \emph{transit-time ultrasonic flow meters} rely on the propagation of ultrasound waves and the computation of their transit time through the moving fluid to estimate the volumetric flow rate \cite{FlowMeasurementHandbook}. Hence, the analysis of the performance of such devices inherently includes the coupling between the acoustic waves dynamics and the underlying fluid flow. \\
Discontinuous Galerkin (DG) methods represent nowadays a noteworthy and widespread alternative to more classical continuous methods for the resolution of partial differential equations (PDEs). Originally proposed in the seventies for hyperbolic equations \cite{ReedHill,Lesaint} and few years later for second-order equations \cite{Babuska,DouglasDupont}, they were disregarded for a long time because of their higher costs due to the larger number of degrees of freedom. The success of these methods came only with  their rediscovery in the nineties thanks to a more in-depth analysis of their properties for conservation \cite{cockburn1989tvb,cockburn1990runge} and diffusion problems \cite{cockburn1998local,Oden1998}. DG methods have indeed proven to be well-suited for polynomial and grid adaptivity \cite{cangiani2014hp} which also allows for highly accurate solutions in many different fields such as the ones stated above \cite{shu2003high,hoermann2018adaptive,nguyen2011high}. Moreover, the decoupling of the adjacent elements facilitates fast parallel computations and leads to some powerful algebraic properties like the efficient inversion of the mass matrix \cite{Hesthaven,chan2018multi}. \\
In this work, we study and develop a DG formulation for ultrasound propagation in a background water flow aiming at simulating the real pressure wave dynamics inside ultrasonic flow meters. While the adoption of discontinuous methods for similar purposes has already been explored in some previous works \cite{luca2014discontinuous,luca2016numerical,kelly2018linear}, in the present research we introduce the Dubiner-Koornwinder spectral basis \cite{koornwinder1975two, Dubiner} and a novel imposition of realistic boundary conditions to improve accuracy and efficiency of the computation but also providing a better representation of the physical phenomena involved. Furthermore, we test the method on a realistic three-dimensional flow metering setting. \\
In Section \ref{Discontinuous Galerkin formulation} we describe the PDE model in terms of the linearized Euler equations for the ultrasound propagation in a moving medium, followed by its space discretization through the discontinuous Galerkin formulation and the time integration using the fourth-order Runge-Kutta scheme. Then, we dedicate Section \ref{Analysis of the acoustic impedance} for a discussion about suitable boundary conditions and we propose a novel strategy for imposing absorbing and resistive walls. To conclude, in Section \ref{Numerical results} the accuracy and good behaviour of the proposed methods are assessed through different tests.

\section{Discontinuous Galerkin formulation}\label{Discontinuous Galerkin formulation}
The linearized Euler equations (LEE) represent one of the most popular models to describe the propagation of sound waves in a fluid \cite{bailly2000,aerospace3040033}. Moreover, their use for ultrasonic water meters applications is already shown in \cite{luca2016numerical}. The model is obtained from the compressible Euler equations decomposing the flow variables into the background flow contribution and the perturbation components caused by the ultrasounds propagation. Since the speed of sound is usually much higher than the fluid motion, it is acceptable to consider a stationary background flow as assumed in the LEE model. Moreover, the use of the inviscid Euler equations is motivated by the low viscosity of the ultrasound waves. Finally, we assume the thermodynamic process to be isentropic, an hypothesis that is often made in acoustics \cite{mechel2004formulas}. 
\subsection{Analytical model for ultrasound propagation}
Let $\Omega\subset \mathbb{R}^d$, where we consider either $d=2$ or $d=3$ dimensions, and let $c\in\mathbb{R}^+$ be the constant speed of sound in the fluid. We define $(\bar{\rho},\bar{p},\bar{\mathbf{u}})$ as the stationary background density, pressure and flow in $\Omega$ respectively. For simplicity, we assume the fluid to be homogeneous and so $\bar{\rho}$ to be constant. Equivalently, let $(\rho,p,\mathbf{u})$ be the perturbation flow components which are instead time-dependent. \\
The linearized Euler equations are therefore:

\begin{equation}\label{LEE model}
    \begin{cases}
    \frac{\partial \rho}{\partial t} + (\bar{\mathbf{u}} \cdot \nabla) \rho + \rho(\nabla \cdot \bar{\mathbf{u}}) + \bar{\rho}(\nabla \cdot \mathbf{u}) = 0, & \Omega \times [0,T],\\
    \frac{\partial \mathbf{u}}{\partial t} + (\bar{\mathbf{u}}\cdot \nabla)\mathbf{u} + (\mathbf{u} \cdot \nabla) \bar{\mathbf{u}} + \frac{c^2}{\bar{\rho}}\nabla \rho - \frac{\rho}{\bar{\rho}^2}\nabla \bar{p} = \mathbf{0}, & \Omega \times [0,T],\\
    \rho = \frac{p}{c^2}, & \Omega \times [0,T],
    \end{cases}
\end{equation}
where $T>0$ is the final time and the energy conservation equation derives from the isentropic assumption.\\
The LEE model is hence a hyperbolic system of $d+2$ equations or $d+1$ if one writes the system only in terms of $\mathbf{u},\rho$ by removing $p$. \\
In contrast with more classical boundary conditions for hyperbolic systems, we opt for the imposition of the boundary flux $\Phi:=\mathbf{u}\cdot \mathbf{n}$ on the entire border $\partial \Omega$, where $\mathbf{n}$ is the outward normal vector with respect to the boundary surface while no condition is imposed for $\rho$. This choice has been recently proven to be well-posed \cite{Galbrun}, and is often adopted in acoustics since it allows to describe conditions that occur in real problems. More details related to the definition of $\Phi$ for the application to realistic simulations will be provided in Section \ref{Analysis of the acoustic impedance}. \\
It is convenient to explicitly suppress the pressure and write Equation (\ref{LEE model}) in divergence form:
\begin{equation}
\label{LEE divergence form}
    \begin{cases}
    \frac{\partial \rho}{\partial t} + \nabla \cdot (\rho\bar{\mathbf{u}}) + \nabla \cdot (\bar{\rho}\mathbf{u}) = 0, & \Omega \times [0,T],\\
    \frac{\partial \mathbf{u}_i}{\partial t} + \nabla \cdot (\mathbf{u}_i \bar{\mathbf{u}}) + \nabla \cdot (\frac{c^2}{\bar{\rho}}\rho \hat{\mathbf{e}}_i) + \mathbf{u} \cdot \nabla \bar{\mathbf{u}}_i   - \frac{\rho}{\bar{\rho}^2}\frac{\partial \bar{p}}{\partial \mathbf{x}_i} - \mathbf{u}_i(\nabla \cdot \bar{\mathbf{u}}) = 0,  & \Omega \times [0,T],\\
    \mathbf{u}\cdot \mathbf{n} = \Phi, &\hspace{-2mm}\partial\Omega \times [0,T],
    \end{cases}
\end{equation}
where the second equation holds for $i=1,\dots,d$ and $\hat{\mathbf{e}}_i$ is the $i$-th standard unitary vector in $\mathbb{R}^d$. \\
The divergence form is now more suitable for the definition of the weak formulation. Multiplying the equations with the test functions $\rho',\mathbf{u}'$, integrating on the domain $\Omega$ and performing integration-by-parts, we obtain
\begin{equation}\label{density weak}
    \int_\Omega \frac{\partial \rho}{\partial t}\rho' + \int_{\partial \Omega} \rho \bar{\mathbf{u}} \cdot \mathbf{n} \rho' - \int_\Omega \rho \bar{\mathbf{u}}\cdot \nabla\rho' + \int_{\partial \Omega} \bar{\rho} \Phi\rho' - \int_\Omega \bar{\rho} \mathbf{u} \cdot \nabla \rho' = 0
\end{equation}
for the density equation and 
\begin{equation}\label{momentum weak}
\begin{gathered}
    \int_\Omega\frac{\partial \mathbf{u}_i}{\partial t} \mathbf{u}_i' + \int_{\partial \Omega}(\mathbf{u}_i \bar{\mathbf{u}})\cdot \mathbf{n} \mathbf{u}_i' - \int_\Omega \mathbf{u}_i \bar{\mathbf{u}}\cdot \nabla \mathbf{u}_i' + \int_{\partial \Omega} \frac{c^2}{\bar{\rho}}\rho\mathbf{n}_i \mathbf{u}_i' - \int_\Omega \frac{c^2}{\bar{\rho}}\rho \frac{\partial \mathbf{u}_i'}{\partial \mathbf{x}_i} + \\ 
    \int_\Omega\mathbf{u} \cdot \nabla \bar{\mathbf{u}}_i \mathbf{u}_i'   - \int_\Omega\frac{\rho}{\bar{\rho}^2}\frac{ \partial \bar{p}}{\partial \mathbf{x}_i}\mathbf{u}_i' - \int_\Omega \mathbf{u}_i(\nabla \cdot \bar{\mathbf{u}})\mathbf{u}_i' = 0
\end{gathered}
\end{equation}
for the momentum equations. We notice again that Equation (\ref{momentum weak}) holds for each $i=1,\dots,d$ and it should not be interpreted as the Einstein notation. Moreover, we note that the assigned boundary flux is weakly imposed in the fourth integral of Equation (\ref{density weak}). The weak imposition of boundary conditions is a well-known strategy and often necessary in the DG framework. We refer the reader to \cite{SCOVAZZI2012117} for the motivation and application of weak boundary conditions to the wave equations.
\subsection{DG space discretization} \label{DG space discretization}
To formally obtain a discretization in space, Equation (\ref{density weak}) and (\ref{momentum weak}) from the original $H^1(\Omega)$ space are restricted to the discretized $V_{h,P}$ space defined as
\begin{equation}\label{V_h}
    V_{h,P}= \{u \in L^2(\Omega): u|_\mathcal{K} \in \mathbb{P}^P(\mathcal{K}) \;\; \forall \mathcal{K} \in \mathcal{T}_h\},
\end{equation}
where $\mathcal{T}_h$ is a conforming triangulation of the domain $\Omega$, $h>0$ indicates the averaged grid size, $\cdot|_\mathcal{K}$ is the restriction operator to the element $\mathcal{K}$ and $\mathbb{P}^P(\mathcal{K})$ is the space of polynomials of order at most $P\in\mathbb{N}_0$. Differently from common FEM spaces, functions in $V_{h,P}$ are in general discontinuous on the edges between adjacent elements but are necessarily continuous inside each element. As already mentioned, this disconnection allows interesting properties, for instance the $P$-adaptivity where $P$ can be chosen differently in each element. On the other hand, DG formulations require the definition of an artificial/numerical flux to be evaluated on the internal edges in order to let the information be shared between the discontinuous elements. We introduce and opt for the Lax-Friedrichs numerical flux which is generally known to be a simple, effective and widespread choice.

\begin{definition}[Discontinuous mean and jump \cite{Hesthaven}] \label{Dg mean and jump}
Let $\mathcal{E}$ be an oriented internal edge with normal vector $\mathbf{n}$. Let $\mathcal{K}^+, \mathcal{K}^- \in \mathcal{T}_h$ be the two elements that share $\mathcal{E}$ and where $\mathcal{K}^+$ is chosen as outward with respect to $\mathbf{n}$. If $u^+,u^-$ are the associated evaluations of a discontinous function $u$, we define
\begin{equation*}
    \{\!\{u\}\!\}:= \frac{1}{2} (u^+ + u^-) 
\end{equation*}
as the discontinuous mean operator and
\begin{equation*}
    [\![u]\!]:= (u^+ - u^-) 
\end{equation*}
the discontinuous jump operator evaluted on $\mathcal{E}$.
\end{definition}

\begin{definition}[Lax-Friedrichs numerical flux \cite{Hesthaven}]\label{DG numerical flux}
Let $\Omega \subset \mathbb{R}^d$ and suppose to solve the following general hyperbolic system
\begin{gather*}
    u: \Omega \rightarrow \mathbb{R},  \\
    \frac{\partial u}{\partial t} + \nabla \cdot F(u) = \mathbf{0},
\end{gather*}
where $F:\mathbb{R}\rightarrow \mathbb{R}^d$ is the flux. The Lax-Friedrichs numerical flux evaluated on an oriented internal edge with normal vector $\mathbf{n}$ is defined as
\begin{equation}\label{LF flux}
    \Upsilon(F(u),u) := \{\!\{F(u) \cdot \mathbf{n}\}\!\} + \frac{\lambda_{max}}{2}[\![u]\!],
\end{equation}
where
\begin{equation*}
    \lambda_{max} := \max_{u} \left| \frac{\partial F}{\partial u} \cdot\mathbf{n}\right|.
\end{equation*}
For multidimensional systems as Equation (\ref{LEE model}) where $\mathbf{u}:\Omega \rightarrow \mathbb{R}^{d+1}$ and $F:\mathbb{R}^{d+1}\rightarrow \mathbb{R}^{d+1,d} $, the same can be generalized through the column-wise application of the scalar product and mean operator.
\end{definition}
In our case, we can infer from Equation (\ref{LEE divergence form}) that $F=[F_1, F_2]$ for $d=2$ is the linear operator defined through the matrices
\begin{equation*}
    \mathcal{F}_1 = \begin{bmatrix}
        \bar{\mathbf{u}}_1 & \bar{\rho} & 0\\
        \frac{c^2}{\bar{\rho}} & \bar{\mathbf{u}_1} & 0 \\
        0 & 0 & \bar{\mathbf{u}_1}
    \end{bmatrix},
    \mathcal{F}_2 = \begin{bmatrix}
        \bar{\mathbf{u}}_2 & 0 & \bar{\rho}\\
        0 & \bar{\mathbf{u}_2} & 0 \\
        \frac{c^2}{\bar{\rho}} & 0 & \bar{\mathbf{u}_2}.
    \end{bmatrix}
\end{equation*}
The Lax-Friedrichs flux is often used because of its simplicity, efficiency and good convergence properties \cite{Hesthaven,brezzi}. While the central flux defined using only the first term in Equation (\ref{LF flux}) is of common usage, it guarantees convergence only in $L^2(\Omega)$ and not in $H^1(\Omega)$ \cite{brezzi}. In practice, this implies that the numerical solution could be quantitatively similar to the exact solution but might have poor regularity. \\
With few calculations, an explicit value for $\lambda_{max}$ can be obtained and the value is approximately $c$ if $|\bar{\mathbf{u}}|\ll c$ (which corresponds to the reasonable hypothesis of subsonic water flow). This is coherent with the physical interpretation of $\lambda_{max}$ that is exactly the maximum speed of the wave. 
Let $\partial \mathcal{T}_h$ be the set of all the oriented internal edges with associated normal $\mathbf{n}$, the DG problem can then be stated: for every $t\in[0,T]$, find $\rho\in V_{h,P}$ and $\mathbf{u}\in V_{h,P}^d$ such that
\begin{equation}\label{dg density weak}
\begin{gathered}
    \sum_{\mathcal{K}\in\mathcal{T}_h}\int_\mathcal{K} \frac{\partial \rho}{\partial t}\rho' + \int_{\partial \Omega} \rho \bar{\mathbf{u}} \cdot \mathbf{n} \rho' - \sum_{\mathcal{K}\in\mathcal{T}_h}\int_\mathcal{K} \rho \bar{\mathbf{u}}\cdot \nabla\rho' + \\ \int_{\partial \Omega} \bar{\rho} \Phi\rho' - \sum_{\mathcal{K}\in\mathcal{T}_h} \bar{\rho} \mathbf{u} \cdot \nabla \rho' + \int_{\partial \mathcal{T}_h} \Upsilon(\rho \bar{\mathbf{u}} + \bar{\rho} \mathbf{u}, \rho) \rho'= 0
\end{gathered}
\end{equation}
and
\begin{equation}\label{dg momentum weak}
\begin{gathered}
    \sum_{\mathcal{K}\in\mathcal{T}_h}\int_\mathcal{K}\frac{\partial \mathbf{u}_i}{\partial t} \mathbf{u}_i' + \int_{\partial \Omega}(\mathbf{u}_i \bar{\mathbf{u}})\cdot \mathbf{n} \mathbf{u}_i' - \sum_{\mathcal{K}\in\mathcal{T}_h}\int_\mathcal{K} \mathbf{u}_i \bar{\mathbf{u}}\cdot \nabla \mathbf{u}_i' + \int_{\partial \Omega} \frac{c^2}{\bar{\rho}}\rho\mathbf{n}_i \mathbf{u}_i' + \\ - \sum_{\mathcal{K}\in\mathcal{T}_h}\int_\mathcal{K} \frac{c^2}{\bar{\rho}}\rho \frac{\partial \mathbf{u}_i'}{\partial \mathbf{x}_i} + 
    \sum_{\mathcal{K}\in\mathcal{T}_h}\int_\mathcal{K}\mathbf{u} \cdot \nabla \bar{\mathbf{u}}_i \mathbf{u}_i'   - \sum_{\mathcal{K}\in\mathcal{T}_h}\int_\mathcal{K}\frac{\rho}{\bar{\rho}^2}\frac{ \partial \bar{p}}{\partial \mathbf{x}_i}\mathbf{u}_i' + \\ - \sum_{\mathcal{K}\in\mathcal{T}_h}\int_\mathcal{K} \mathbf{u}_i(\nabla \cdot \bar{\mathbf{u}})\mathbf{u}_i'+ \int_{\partial \mathcal{T}_h} \Upsilon(\mathbf{u}_i\bar{\mathbf{u}} + \frac{c^2}{\bar{\rho}}\rho\hat{\mathbf{e}}_i, \mathbf{u}_i)\mathbf{v}_i' = 0
\end{gathered}
\end{equation}
for every $i=1,\cdots,d$, $\rho'\in V_{h,P}$ and $\mathbf{u}'\in V_{h,P}^d$. \\
Following the classical DG theory, volume integrals are performed element-wise so that the arguments are continuous and integrable disregarding the internal jumps. Notice that the integration of the numerical flux yields no ambiguity in the evaluations of the discontinuous functions on the edges. Moreover, the integrals on the boundary $\partial \Omega$ evaluate the functions on the internal border side. \\
Let $\{\hat{\varphi}_j\}_{j=1,\dots,N_{h,P}}$ be a set of basis functions for $V_{h,P}$ where $N_{h,P}\in\mathbb{N}$ is the dimension of the discretized space. The explicit definition of such basis is postponed to Section \ref{Choice for the spectral basis} as well as the quadrature formulae that are used to compute the integrals in Equation (\ref{dg density weak}) and (\ref{dg momentum weak}). Furthermore, let $\mathbf{u}_{h,P}\in \mathbb{R}^{(1+d)N_{h,P}}$ be the vector containing the Fourier coefficients of the flow unknowns $(\rho,\mathbf{u})$ with respect to $\{\hat{\varphi}_j\}_{j=1,\dots,N_{h,P}}$. It is therefore convenient to create $(d+1)$ copies of the basis functions which we define as $\varphi_j := \hat{\varphi}_{(j \mod N_{h,P})}$. Equation (\ref{dg density weak}) and (\ref{dg momentum weak}) can then be rewritten in the following compact algebraic form, that will be useful for the upcoming sections:
\begin{equation}\label{compact form}
    \mathcal{M}\frac{\partial\mathbf{u}_{h,P}}{\partial t} + \mathcal{S}(t)\mathbf{u}_{h,P} = \mathbf{0},
\end{equation}
where $\mathcal{M}_{j,k}:=<\varphi_j,\varphi_k>_{L^2(\Omega)}$ is the mass matrix and $\mathcal{S}(t)$ is the matrix associated with all the remaining terms in Equation (\ref{dg density weak}) and (\ref{dg momentum weak}). These terms are indeed linear with respect to $(\rho,\mathbf{u})$ and the dependence on $t$ is due to the imposition of the boundary flux $\Phi(t)$. It is common to store only the local mass matrix applied to the reference element since it can be quickly reconstructed for each affinely mapped element. On the contrary, this strategy is evidently not applicable to the $\mathcal{S}$ matrix except for constant background flows. A low-storage method is still possible by assembling each local $\mathcal{S}$ at every time step but this approach significantly undermines the efficiency of the scheme. In conclusion, a global assembly is still considered due to its considerable performance despite the burden in terms of memory.

\subsection{Efficient explicit-time discretization}\label{efficient explicit time discretization}
In this section the original problem in Equation (\ref{LEE model}) is fully discretized with a explicit-time numerical integration. Let us introduce the partitioning of the time interval $[0,T]$ as $0=t_0<t_1<\dots<t_M=T$ where $\Delta t = T/M$ is the uniform difference between two consecutive time steps and $M+1\in \mathbb{N}$ is the total number of time steps. Moreover, let $\mathbf{u}^n_{h,p}$ be the numerical evaluation of the solution vector at the $n$-th time step, where $0\le n\le M$. $\mathbf{u}^0_{h,p}$ is inherently the projection on $V_{h,P}$ of the problem initial condition. \\
We require an accurate and stable time discretization method that allows to efficiently perform ultrasound simulations with thousands of time steps. Runge-Kutta discontinuous Galerkin (RKDG) methods have been consolidated as a class of powerful methods for the full discretization of convection-dominated problems. Proposed and studied in a series of works \cite{cockburn1989tvb,cockburn1990runge}, they have been proven to be at the same time fast and accurate. Fourth-order Runge-Kutta (RK4) is hence chosen and its application to Equation (\ref{compact form}) performs the following operations at each time step:
\begin{equation}\label{RK4}
\begin{cases}
        \frac{1}{\Delta t}\mathcal{M}\mathbf{z}_{1} + \mathcal{S}(t_n)\mathbf{u}_{h,P}^{n}= \mathbf{0}, & \\
        \frac{1}{\Delta t}\mathcal{M}\mathbf{z}_{2} + \mathcal{S}(t_n+\Delta t/2)(\mathbf{u}_{h,P}^{n}+\frac{1}{2}\mathbf{z}_{1}) = \mathbf{0}, \\
        \frac{1}{\Delta t}\mathcal{M}\mathbf{z}_3+ \mathcal{S}(t_n+\Delta t/2)(\mathbf{u}_{h,P}^{n}+\frac{1}{2}\mathbf{z}_2) = \mathbf{0}, & \\
        \frac{1}{\Delta t}\mathcal{M}\mathbf{z}_4 + \mathcal{S}(t_{n+1})(\mathbf{u}_{h,P}^{n}+\mathbf{z}_3) = \mathbf{0}, \\
        \mathbf{u}_{h,P}^{n+1} = \mathbf{u}_{h,P}^{n} + \frac{1}{6}\mathbf{z}_1 + \frac{1}{3}\mathbf{z}_2 + \frac{1}{3}\mathbf{z}_3 + \frac{1}{6}\mathbf{z}_4.
\end{cases}
\end{equation}
In Equation (\ref{RK4}), the vectors $\mathbf{z}_1,\mathbf{z}_2,\mathbf{z}_3,\mathbf{z}_4$ have the role of intermediate step solutions and allow for the fourth-order accuracy. Many different alternatives to RK4 are possible, for instance the choice of a different order or the adoption of Strong-Stability-Preserving (SSP) Runge-Kutta methods \cite{ssp,ssp2e}. However, due to the linearity of the LEE model in Equation (\ref{LEE model}) and the accuracy order that is usually desired for such simulations, RK4 turns out to be a good compromise of accuracy and simplicity. \\
Stability is achieved by satisfying the well-known Courant-Friedrichs-Lewy (CFL) condition
\begin{equation*}
    v_{CFL} \frac{\Delta t}{h} \le k_{CFL},
\end{equation*}
where $k_{CFL}\in \mathbb{R}^+$ is a characteristic parameter of the adopted RKDG scheme and $v_{CFL}\in\mathbb{R}^+$ is the maximum wave speed. The latter can be computed from the eigenvalues of the problem flux $F$, thus it coincides with $\lambda_{max}$ whose value is approximately $c$ when the background flow is subsonic, coherently with the physical intuition. In conclusion, given $k_{CFL}$, $M$ must be large enough to satisfy the CFL condition with $v_{CFL}=c$.\\  
We remind that the time update of $\mathcal{S}(t)$ can be performed by updating only the surface integral associated to the boundary flux $\Phi$. Moreover, the mass matrix needs to be inverted only once and this can be achieved efficiently element-wise. Indeed, since each DG basis function has support in a single element, the mass matrix has an element-block structure. This, coupled with the performances of the RKDG method, leads to a very efficient time stepping strategy.

\subsection{Spectral basis and quadrature formulae} \label{Choice for the spectral basis}
While the above consideration hold for a wide class of DG methods (e.g., when adopting the nodal discontinuous Lagrangian basis), in this work we choose to adopt the modal Dubiner-Koornwinder basis \cite{koornwinder1975two, Dubiner}. These spectral polynomials are constructed from a transformation that collapses quadrilaterals on simplices to obtain a $L^2$-orthogonal basis on triangles and tetrahedra which is formed by generalized tensor products of Jacobi polynomials. Their explicit definitions are reported in \ref{Dubiner-Koornwinder polynomials}. \\
The $L^2$ basis orthogonality leads to a diagonal mass matrix which considerably reduces time and memory requirements. In particular, if $N_P\in \mathbb{N}$ is the number of degrees of freedom per element, the number of non-zero elements with nodal DG bases is only $\mathcal{O}(N_{h,P} N_{P})$ but it can be even reduced to $\mathcal{O}(N_{h,P}$) with a diagonal mass matrix. This discrepancy also affects the performance during the pre-inversion of the matrix and numerous matrix-vector products. While the former operation is executed only once, mass matrix products are performed at each time step. Being $N_P=(P+1)(P+2)/2$ in $d=2$ dimensions and $N_P=(P+1)(P+2)(P+3)/6$ in 3 dimensions, it is clear that substantial differences appear when performing high-order discretizations (or even only $P\ge 3$). These expected performances improvement are verified in Section \ref{spectral basis performance} where the ultrasounds propagation is simulated with order $P=4$. \\
To conclude this section, a proper quadrature formula is defined for the integrals in the weak formulation of Equation (\ref{dg density weak}) and (\ref{dg momentum weak}) as well as for the computation of the $L^2$ Dubiner projection coefficients. The Gauss-Legendre quadrature nodes are known to work well with high-order discretizations using spectral bases and they represent therefore the reference choice in one dimension \cite{Hesthaven}. Different possible options are available to extend in 2 and 3 dimensions and we opted for the so-called symmetrical Gauss formulae, proposed by \cite{quadrature} and provided by the XLiFE++ library \cite{kielbasiewicz2017xlife++}. We note that internal quadrature formulae are preferred because they allow to avoid some complications due to the singularities of the Dubiner functions on the borders. 
\section{Application of acoustic impedance for absorbing and resistive boundaries}\label{Analysis of the acoustic impedance}
As stated in Section \ref{Discontinuous Galerkin formulation}, our boundary conditions prescribe the assignment of the boundary flux $\Phi=\mathbf{u}\cdot\mathbf{n}$ on the entire boundary. While perfectly reflecting walls and symmetry planes can be achieved by simply assigning $\Phi=0$, the simulation of absorbing layers (i.e., walls that capture incoming waves without reflections) is more complicated, especially in higher spatial dimensions. We note that absorbing layers are crucial when the real domain is infinite or very large and simulations are instead performed in a smaller region as a finite section in an ultrasonic flow meter pipe. For such reasons, different methods have been proposed in literature as the so-called perfectly matched layers (PML) where the wave decays following a reaction-diffusion model in an external and artificial region. However, PMLs have been proven as not always accurate (e.g., in \cite{qi1998evaluation}) and require an additional cost for the problem resolution in an extra domain. What we propose here is a novel, truthful and realistic method that is based on the imposition of the acoustic impedance. The following definitions are readapted from \cite[Section 6.3]{Pierce}.
\begin{definition}[Acoustic impedance of a surface]\label{acoustic impedance of a surface}
    Let an acoustic wave be represented in the complex exponential formalism ($Ae^{i(\mathbf{k}\cdot \mathbf{x}) - \omega t}$). We define the acoustic impedance with respect to a surface as the ratio
    \begin{equation*}
        \mathcal{Z}(\mathbf{x},t):= \frac{p}{\mathbf{u}\cdot \mathbf{n}} \in \mathbb{C}
    \end{equation*}
    where $\mathbf{n}$ is the normal vector to the surface that is outward with respect to the incoming wave.
\end{definition}

\begin{proposition}[Acoustic impedance on the interface between two homogeneous mediums]\label{acoustic impedance on the interface}
Let $(\bar{\rho},c)$ and $(\rho_w, c_w)$ be the density and speed of sound in two mediums. Assume the width of the second medium to be infinite. The impedance on the interface with respect to the waves coming from the first medium is
\begin{equation*}
    \mathcal{Z}_w = \frac{\rho_w c_w}{\cos(\theta_p)}
\end{equation*}
where $\theta_p$ is the penetration angle that can be obtained from the incidence angle using the acoustic Snell's law:
\begin{equation*}
    \frac{\sin(\theta)}{c} = \frac{\sin(\theta_p)}{c_w}.
\end{equation*}
\end{proposition}
From Proposition \ref{acoustic impedance on the interface} we therefore find a proper boundary flux imposition for different wall materials. If the wall is assumed to be thick and rigid enough, which is needed in practice to neglect resonance effects (\cite[Chapter 11]{auld1990acoustic}), we have a condition of realistic/dissipative wall. With few calculations, the boundary flux can be written as
\begin{equation}\label{realistic wall}
\Phi = \sqrt{\max\{c^2-c_w^2 + c_w^2 \cos^2(\theta),0\}} \frac{\rho c}{\rho_w c_w},
\end{equation}
where $\theta$ is the angle of incidence of the sound wave. Note that the argument of the square root is zero when the critical angle of the Snell's law is achieved. When this happens, the impedance wall behaves as a perfectly reflecting wall (i.e., $\Phi=0$ or equivalently $\mathcal{Z}_w=\infty$). \\
The condition for an absorbing layer can be obtained as a specific case of Equation (\ref{realistic wall}), i.e., when the wall material is the same as the conductive medium. The absorbing condition will therefore read as
\begin{equation}\label{absorbing wall}
\Phi = \lvert\cos(\theta)\rvert \frac{\rho c}{\bar{\rho}}.
\end{equation}
Some difficulties arise to determine or estimate the wave propagation angle $\theta$. Indeed, it is well-known that imposing the impedance in $d>1$ dimensions can be considerably more complex and that most of the strategies proposed involve frequency domain analysis. In the presence of absorbing layers, we opt instead for a novel strategy that consists in the estimation of the angle of incidence directly from the incoming velocity field in the time domain. In an explicit setting as the RK discretization from Section \ref{efficient explicit time discretization}, if $\eta:=\lvert \cos(\theta)\rvert$ is the direction coefficient, we propose the following algorithm:
\begin{equation} \label{filtering}
    \begin{cases}
        \hat{\mathbf{u}}^{n+1} = (1-\alpha_M)\hat{\mathbf{u}}^{n} + (\alpha_M) \sgn{\mathbf{u}^n\cdot \mathbf{n}}\mathbf{u}^n, \\
        \eta^{n+1} = \frac{\lvert \hat{\mathbf{u}}^{n+1}\cdot \mathbf{n}\rvert}{\|\hat{\mathbf{u}}^{n+1}\|}.
    \end{cases}
\end{equation}
Here, $\alpha_M \in (0,1)$ is a memory coefficient and $\hat{\mathbf{u}}$ takes the role of filtered velocity field on the border that can be initialized with a very small value (such as $ 10^{-10}$ despite it depends on the problem under consideration). The filtering operation provides a stable value for the propagation direction also in proximity of the zeroes of the waves. The sign function is instead needed to ensure monotonicity to the filtering operation. \\
Equation (\ref{filtering}) can be used in the absorbing impedance condition from Equation (\ref{absorbing wall}) while it is not meant to be used in the dissipative wall condition in Equation (\ref{realistic wall}). In the latter case the sound reflection does not allow a simple formula for the estimate of the incoming wave direction. Although, other solutions are certainly possible when the ultrasound trajectory is already known or can be estimated. Under this hypothesis, the angle can be provided a priori and, as previously mentioned, dissipative walls can be imposed as perfectly reflecting for high angles of incidence.

\section{Numerical results}\label{Numerical results}
This section is devoted to numerical tests assessing the behaviour and performance of the above proposed methods. The code has been implemented using \texttt{FreeFem++ 4.12} \cite{freefem} adding the Dubiner basis and high-order quadrature formulae as plugins. Moreover, the Message Passing Interface (MPI) has been adopted to parallelize the matrix assembly and the matrix-vector products through a domain decomposition method; details are provided in Section \ref{Parallel calculations}. In the following tests we always consider $c=1481$ m/s as the speed of sound in water and $\Bar{\rho}=997$ kg/m$^3$ as the density of water. All the calculations are performed on CPUs \texttt{Intel Xeon E5-2699 v3 @ 2.30 GHz}. Moreover, Gmsh 4.8 \cite{geuzaine2009gmsh} is employed to generate the unstructured grid $\mathcal{T}_h$. We will associate the grid length $h>0$ with the parameter provided to the software that specifies the level of refinement on the boundary.

\subsection{Convergence analysis}\label{convergence analysis}
\begin{figure}[ht]
 \begin{subfigure}{0.5\textwidth}
 \centering
 \begin{tikzpicture}[scale=0.75]
\begin{loglogaxis}[
    xlabel={h (m)},
    ylabel={$H^1$ error},
    xmin=0.02/250, xmax=0.02/8,
    ymin=4e-5, ymax=1e-2,
    legend pos=south east,
    ymajorgrids=true,
    grid style=dashed,
]

\addplot[
    color=blue,
    mark=*,
    ]
    coordinates {
    (0.00015384615384615385,0.005889000033977141)(0.00013333333333333334,0.004987000017744986)(0.00011764705882352942,0.004345000009591218)(0.00010526315789473685,0.0038610000058532453)(9.523809523809524e-05,0.003489000003518901)(8.695652173913044e-05,0.0032290000021891006)
    };

\addplot[
    color=red,
    mark=x,
    ]
    coordinates {
    (0.0004,0.003088000013523478)(0.0003333333333333333,0.0021220000036875617)(0.00028571428571428574,0.0015660000012396346)(0.00025,0.0012570000005302229)(0.00022222222222222223,0.000962100000245397)(0.00018181818181818183,0.0006830000000898729)(0.00015384615384615385,0.00047200000003937864)
    };

\addplot[
    color=green,
    mark=square*,
    ]
    coordinates {
    (0.0008,0.003587000030870415)(0.0006666666666666666,0.0020960000055623308)(0.0005714285714285715,0.0015780000015695008)(0.0005,0.0010200000005967217)(0.0004,0.0005246000001534846)(0.0003333333333333333,0.0003202000000573826)(0.00028571428571428574,0.00021010000003200116)(0.00025,0.00014650000001964234)
    };

\addplot[
    color=cyan,
    mark=triangle*,
    ]
    coordinates {
    (0.001,0.0020690000076999155)(0.0008,0.0010590000010991796)(0.0006666666666666666,0.0005000000004687225)(0.0005,0.00018220000015358224)(0.0004,8.634000006631606e-05)
    };

\addplot[
    color=violet,
    mark=|,
    ]
    coordinates {
    (0.0013333333333333333,0.0024030000089157303)(0.001,0.0007700000013878005)(0.0008,0.00029560000067794384)(0.0006666666666666666,0.00016070000032244056)
    };
    \legend{$V_{h,1}$, $V_{h,2}$, $V_{h,3}$, $V_{h,4}$, $V_{h,5}$}
    
    \logLogSlopeTriangle{0.15}{0.08}{0.75}{1}{black};
    \logLogSlopeTriangle{0.3}{0.08}{0.4}{2}{black};
    \logLogSlopeTriangle{0.4}{0.08}{0.12}{3}{black};
    \logLogSlopeTriangle{0.55}{0.08}{0.08}{4}{black};
    \logLogSlopeTriangle{0.85}{0.08}{0.55}{5}{black};
    
\end{loglogaxis}
\end{tikzpicture}
\caption{Grid refinement convergence.}
\label{fig:mono21}
\end{subfigure}
\begin{subfigure}{0.5\textwidth}
\centering 
\begin{tikzpicture}[scale=0.75]
\begin{loglogaxis}[
    xlabel={Simulation time (s)},
    ylabel={$H^1$ error},
    xmin=8, xmax=500,
    ymin=4e-5, ymax=1e-2,
    legend pos=south west,
    ymajorgrids=true,
    grid style=dashed,
]

\addplot[
    color=blue,
    mark=*,
    ]
    coordinates {
    (76.63,0.005889000033977141)(119.1,0.004987000017744986)(175.0,0.004345000009591218)(244.0,0.0038610000058532453)(332.5,0.003489000003518901)(434.9,0.0032290000021891006)
    };

\addplot[
    color=red,
    mark=x,
    ]
    coordinates {
    (22.82,0.003088000013523478)(39.47,0.0021220000036875617)(63.48,0.0015660000012396346)(95.54,0.0012570000005302229)(133.3,0.000962100000245397)(248.9,0.0006830000000898729)(405.7,0.00047200000003937864)
    };

\addplot[
    color=green,
    mark=square*,
    ]
    coordinates {
    (11.08,0.003587000030870415)(17.76,0.0020960000055623308)(27.85,0.0015780000015695008)(42.58,0.0010200000005967217)(81.76,0.0005246000001534846)(140.2,0.0003202000000573826)(222.9,0.00021010000003200116)(331.3,0.00014650000001964234)
    };

\addplot[
    color=cyan,
    mark=triangle*,
    ]
    coordinates {
    (18.13,0.0020690000076999155)(32.27,0.0010590000010991796)(54.05,0.0005000000004687225)(125.6,0.00018220000015358224)(243.4,8.634000006631606e-05)
    };
    
\addplot[
    color=violet,
    mark=|,
    ]
    coordinates {
    (20.33,0.0024030000089157303)(42.44,0.0007700000013878005)(75.23,0.00029560000067794384)(125.5,0.00016070000032244056)
    };
    \legend{$V_{h,1}$, $V_{h,2}$, $V_{h,3}$, $V_{h,4}$, $V_{h,5}$}
    
\end{loglogaxis}
\end{tikzpicture}
    \caption{Efficiency in time.}
\end{subfigure}
\caption{Convergence and efficiency study of the RKDG methods for different polynomial orders. a) All the methods have a correct convergence behaviour after grid refinements even if fourth and fifth orders are not fully achieved. b) Low-order methods are less accurate for the same calculation time and the $P\ge 3$ choice is recommended.  }
\label{fig:convergence}
\end{figure}
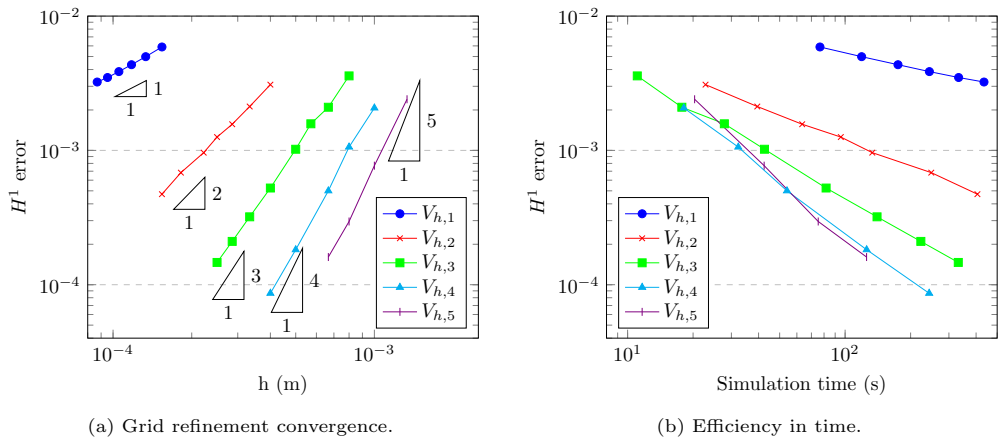

We consider a simple two-dimensional test where the exact analytical solution is known. Therefore, we can assess the convergence of the proposed methods and perform an accuracy comparison for different polynomial orders. \\
Let the background flow as well as the initial ultrasound flow be at rest ($\Bar{\mathbf{u}}=\mathbf{0}$ and $\Bar{p}$ constant). This choice is motivated by the fact that the flow convection is very slow with respect to the wave dynamics and therefore it can be correctly simulated at any order. We define the domain as a square $\Omega = \{0\text{ m}\le x \le 0.02\text{ m}, 0\text{ m} \le y \le 0.02\text{ m}\}$ and the source inlet, to be applied only on the left edge, as 
\begin{equation}\label{test_inlet}
\Phi(x,y,t) := 10^{-3}\sin(2\pi f t) \mathcal{I}_{[0,5/f]}(t) \text{ m/s},
\end{equation}
where $f=10^6$ Hz is the source frequency and $\mathcal{I}$ is the indicator function. The exact solution of the horizontal velocity is therefore known to be 
\begin{equation}\label{uex}
    u_{ex}(x,y,t):=-10^{-3}\sin\left(2\pi f \left(t-\frac{x}{c}\right)\right) \mathcal{I}_{\left[c\left(t-\frac{5}{f}\right),ct\right]}(x) \text{ m/s},
\end{equation}
whereas the vertical velocity is uniformly zero. In addition, one can utilize Equation (\ref{LEE model}) to obtain the simple Cauchy problem 
\begin{equation*}
\begin{cases}
    \frac{\partial \rho_{ex}}{\partial t} = - \bar{\rho} \frac{\partial u_{ex}}{\partial x}, & \forall (x,y) \in \Omega, \forall t \in (0,T],  \\
    \rho_{ex}(t=0) = 0, & \forall (x,y) \in \Omega,
\end{cases}
\end{equation*}
which, coupled with Equation (\ref{uex}), yields $\rho_{ex}=-\bar{\rho}u_{ex}/c$. We set a final time $T=10^{-5}$ s and the largest time step such that the CFL condition for stability is achieved. It has been experimentally verified that this condition alone entails a negligible time discretization error with respect to the total error. Moreover, the CFL coefficient $k_{CFL}$ has been confirmed to follow a trend in accordance to the $1/(2P+1)$ bound from \cite{cockburn1989tvb}. \\
Results are illustrated in Figure \ref{fig:convergence} where the standard $H^1(\Omega)$ error has been computed with respect to $u_{ex}$ through single-core simulations. We remind that the $H^1(\Omega)$ error is defined as $\|u-u_{ex}\|_{H^1(\Omega)}^2:=\|u-u_{ex}\|_{L^2(\Omega)}^2 + \| |\nabla u-\nabla u_{ex}| \|^2_{L^2(\Omega)}$ even if the first contribution is almost negligible due to the fast variation of the functions considered. 
We observe that all the methods exhibit the correct convergence trend $O(h^P)$ even if higher rates (i.e., for $P=4,5$) are not fully achieved. This is presumably due to the fact that the CFL condition is more strict for greater orders and many time steps lead to some dissipation from the Lax-Friedrichs flux.  On the other hand, the plot on the right shows how low-order methods represent the worst trade-off between simulation time and accuracy while higher orders ($P\gtrsim 3$) still represent the best choice for high-fidelity simulations of ultrasounds.

\subsection{Spectral basis performance}\label{spectral basis performance}
To assess the benefits from the choice of a spectral method (Section \ref{Choice for the spectral basis}), we devote this section to a comparison between the Lagrangian and Dubiner bases in terms of memory and time costs. 
We run in a single core the same test case as in Section \ref{convergence analysis} with $P=4$ and $h=6.67\cdot10^{-4}$ m. Results are shown in Table \ref{time table} where the nodal Lagrangian basis is defined as in \cite{Hesthaven}. \\
Although the use of complex spectral functions entails a slightly longer time to assemble the DG matrices, this is minimal compared to the improvement in the inversion of the mass matrix and the total simulation time. As already mentioned in Section \ref{Choice for the spectral basis}, the mass matrix using nodal DG bases is block diagonal where each block corresponds to an element. A naive inversion would lead to $O(N_{h,P}N_P^2)$ operations but it can be even decreased to $O(N_{h,P}N_P)$ by only inverting the local mass matrix associated to the reference element and scaling by the Jacobian of the affine map. Since the diagonal spectral matrix yields a $O(N_{h,P})$ inversion cost, the ratio between the values in the second line is as expected approximately $N_4=15$ in two dimensions. We remark again that spectral bases can be leveraged to improve the cost of the pre-inversion operation which, despite being performed only once, it could correspond to a substantial burden when the number of time steps is limited. \\  
In addition, multiple matrix-vector products have to be performed at each time step using the inverse of the mass matrix. Therefore, a diagonal mass matrix speeds up the total simulation time and reduce the memory requirements. However, a comparison between the values in the two columns is in this case less explicable since these both include the considerable cost to save and perform operations with the operator $\mathcal{S}$.

\begin{table}[ht]
\begin{center}
\begin{tabular}{ l | c | c }
   & Lagrangian basis & Dubiner basis \\
   \midrule
  Matrices assembly time & 43.57 s & 46.59 s \\
  Mass matrix inversion time & 30.18 s & 1.69 s \\
  Simulation time* & 604 s & 495 s \\
  RES memory & 1.31 GB & 0.84 GB \\
  \bottomrule
  \multicolumn{3}{l}{\footnotesize *execution of all the time steps after the assembly and inversion preprocessing phases.}\\
\end{tabular}
\end{center}
\caption{Performance results between the two DG bases with polynomial order $P=4$.}\label{time table}
\end{table}

\subsection{Totally absorbing wall tests}\label{totally absorbing wall tests}
\begin{figure}[ht]
\centering
\begin{subfigure}{0.20\textwidth}
  \centering
  \includegraphics[width=1\linewidth,height=2\linewidth]{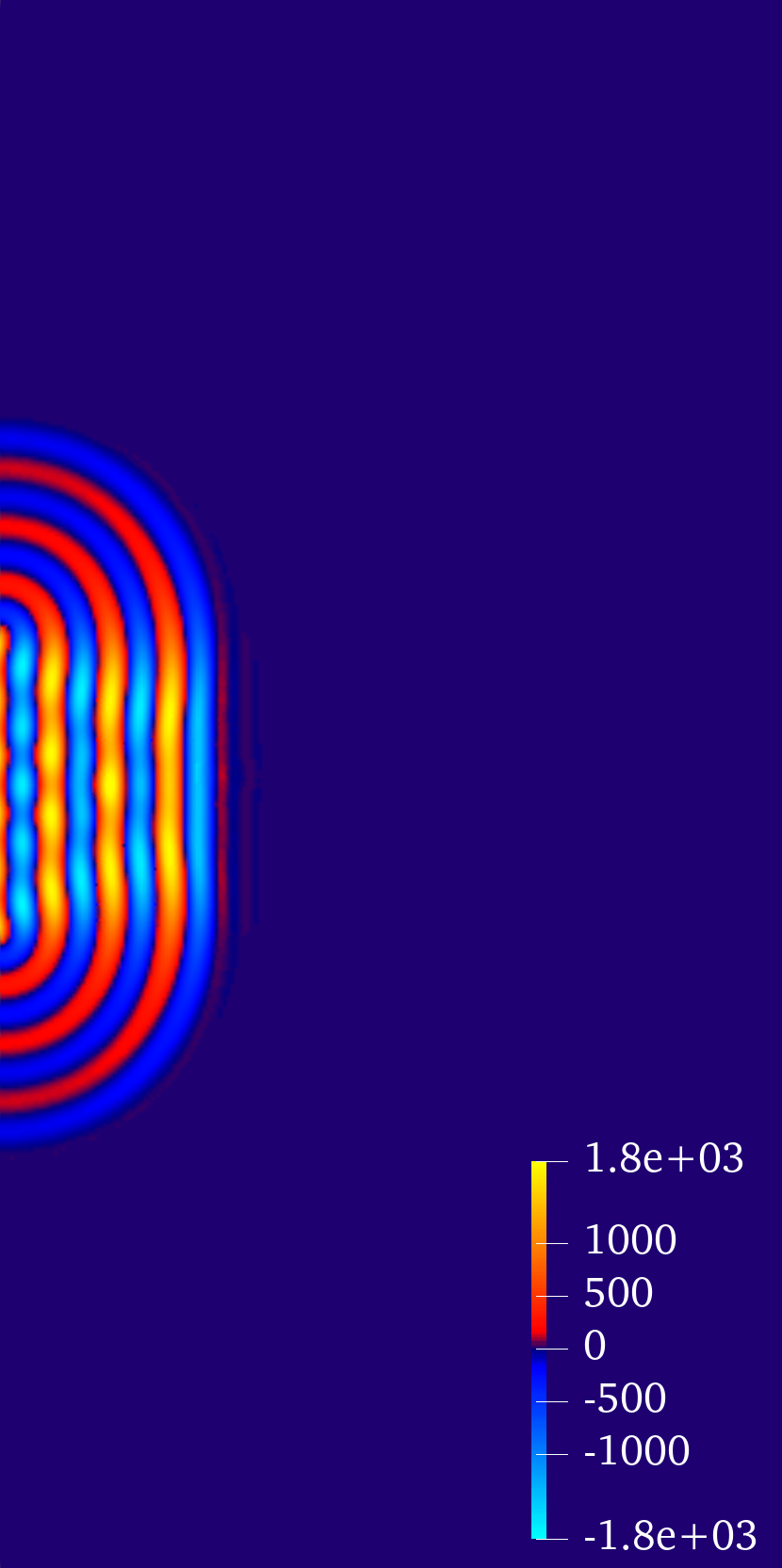}
  \caption{$t=3\cdot 10^{-6}$ s}
\end{subfigure}%
\hspace{0.01\textwidth}
\begin{subfigure}{.20\textwidth}
  \centering
  \includegraphics[width=1\linewidth,height=2\linewidth]{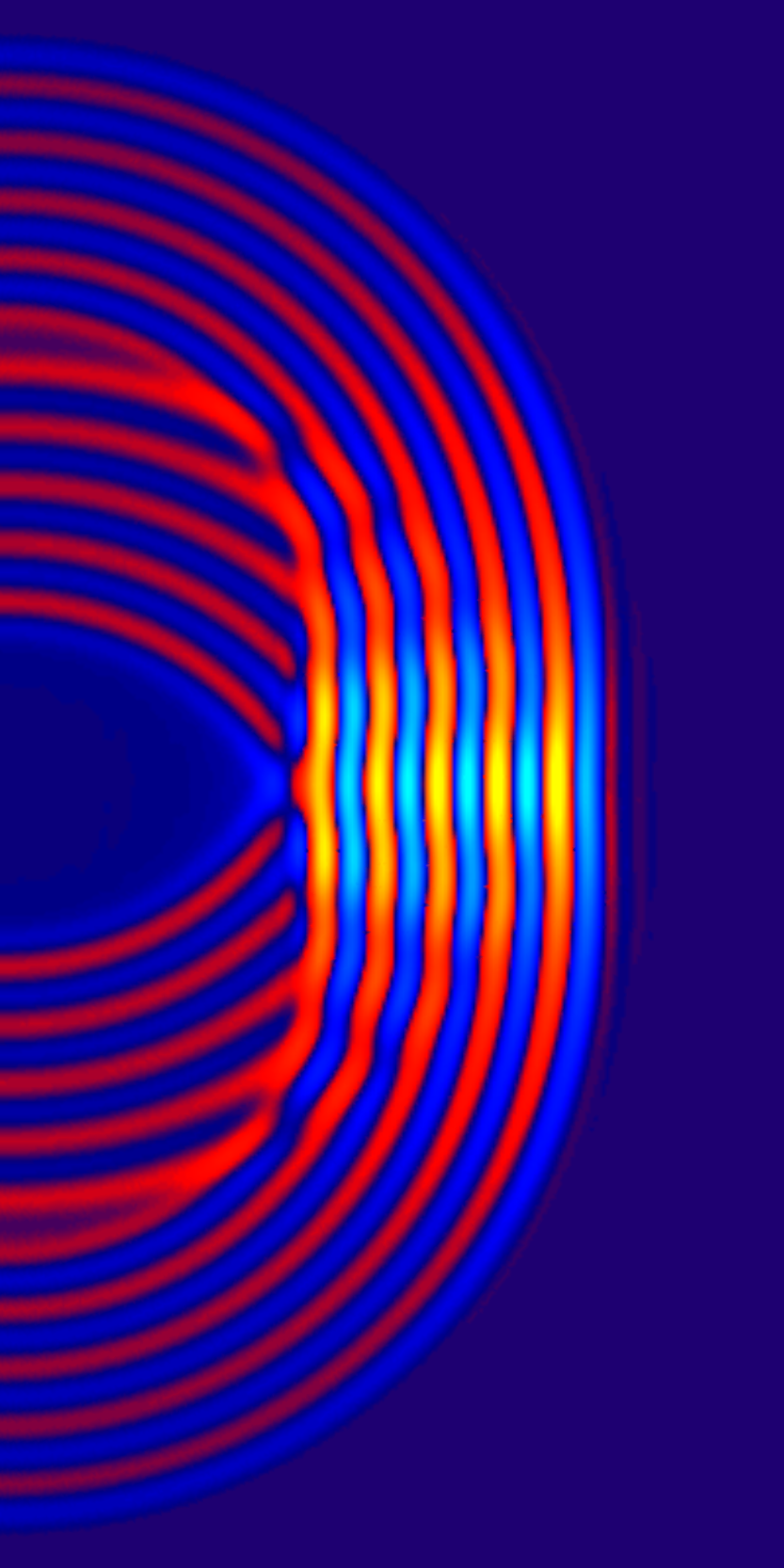}
  \caption{$t=1\cdot 10^{-5}$ s}
\end{subfigure}%
\hspace{0.01\textwidth}
\begin{subfigure}{.20\textwidth}
  \centering
  \includegraphics[width=1\linewidth,height=2\linewidth]{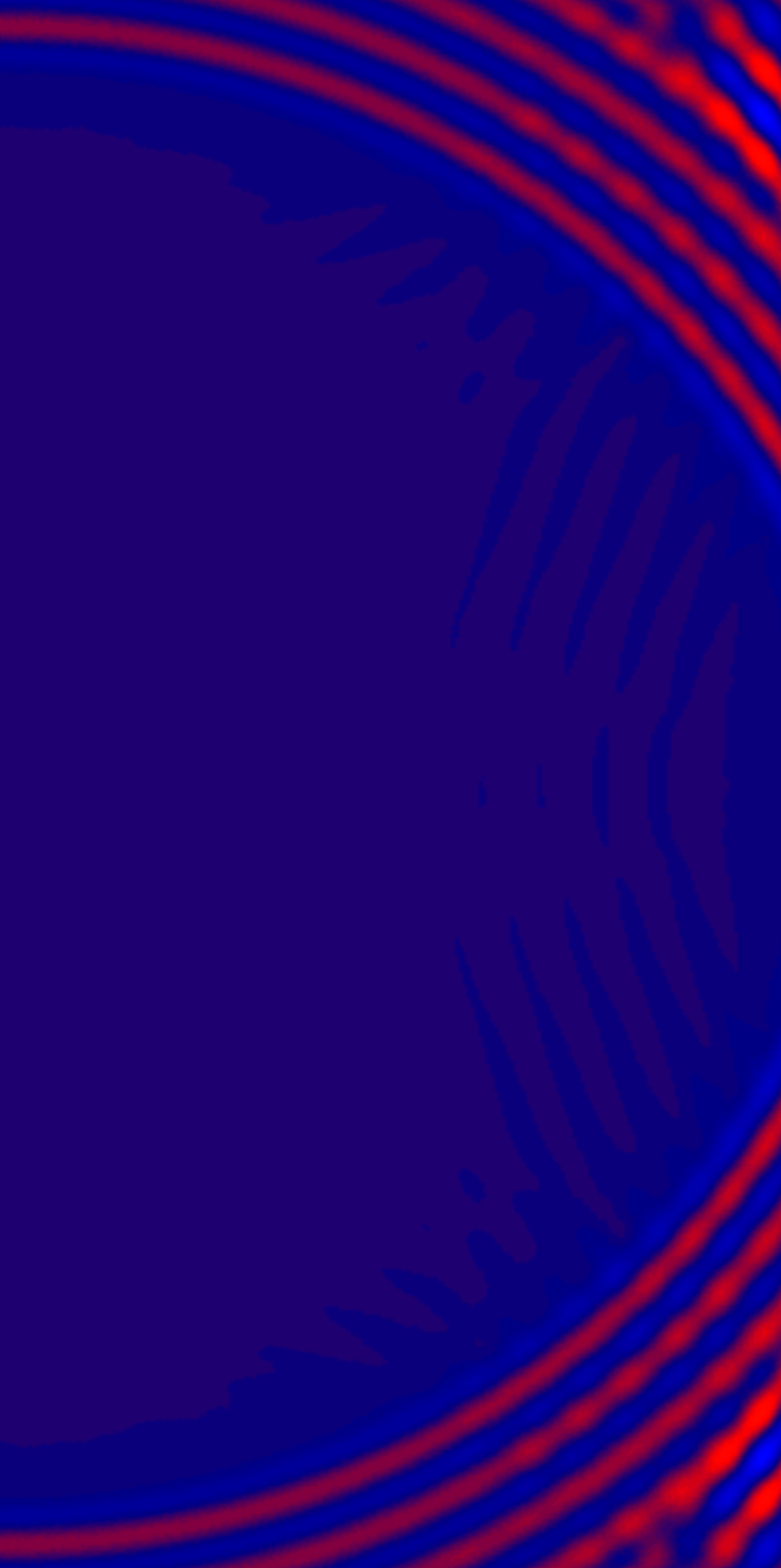}
  \caption{$t=2\cdot 10^{-5}$ s}
\end{subfigure}%
\hspace{0.01\textwidth}
\begin{subfigure}{.20\textwidth}
  \centering
  \includegraphics[width=1\linewidth,height=2\linewidth]{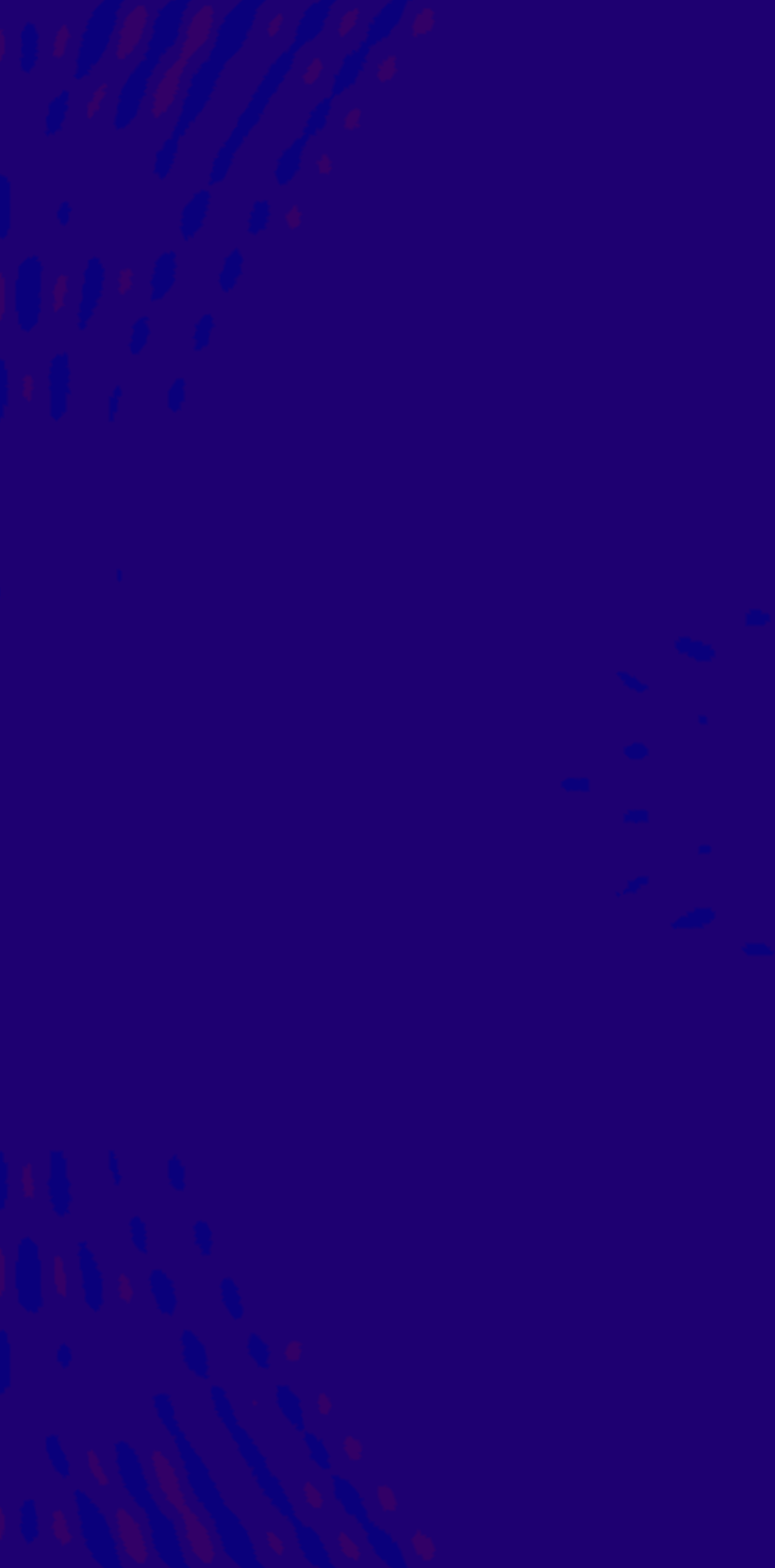}
  \caption{$t=3\cdot 10^{-5}$ s}
\end{subfigure}%
\caption{Snapshots of the test case proposed in Section \ref{totally absorbing wall tests} where the same color ranges are used for the pressure $p$ (Pa) in the four pictures. As expected, the wave is not dissipated nor reflected on the walls.}
\label{fig:absorbing}
\end{figure}
In this section, we test the efficacy of the proposed algorithm for the imposition of the absorbing layers (Equation (\ref{filtering})). \\
The domain is rectangular: $\Omega = \{0\text{ m}\le x \le 0.02\text{ m}, 0\text{ m} \le y \le 0.04\text{ m}\}$ while the background is defined as a standard pipe Poiseuille flow:
\begin{equation}\label{poiseuille}
\begin{cases}
    \bar{u}(x,y) &= \frac{G}{4\mu}  \left(R^2 - (y-R)^2\right), \\
    \Bar{v}(x,y) &= 0, \\
    \Bar{p}(x,y) &= p_0 - Gx,
\end{cases}
\end{equation}

\begin{figure}[ht]
    \centering
    \includegraphics[width=0.7\textwidth]{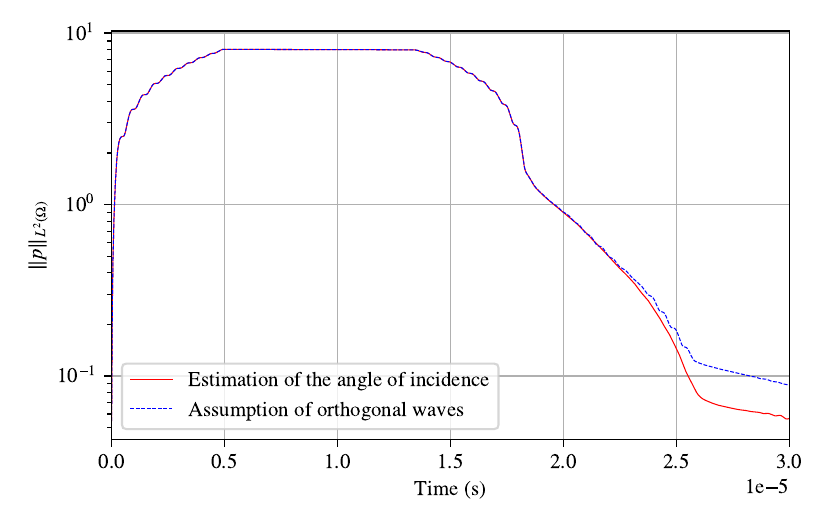}
    \caption{$L^2$ norm of the perturbation pressure. The red line corresponds to the algorithm in Equation (\ref{filtering}). The dotted blue line is obtained instead when the absorbing condition in Equation (\ref{absorbing wall}) is used with $\theta=0$, i.e., assuming that the incident waves are orthogonal to the wall.}
    \label{fig:decay}
\end{figure}
where $R=0.02$ m is the radius of the pipe, in this case half of the domain height, $\mu=1.0016\cdot 10^{-3}$ Pa$\cdot $s is the dynamic viscosity of water, and $G>0$ is such that the velocity in the center of the domain is $u_{max}=20$ m/s. The value of the pressure offset $p_0$ is irrelevant since the LEE equations include only the derivatives of the background pressure and, thus, we can arbitrarily assign $p_0=0$ Pa. \\
Acoustic waves are generated at the left edge following the same profile as in Equation (\ref{test_inlet}) but with support only in $y\in [0.016\text{ m},0.024\text{ m}]$. To test the impedance condition, all the edges except the inlet are set as absorbing with $\alpha_M=0.01$. Motivated by the previous convergence tests, we choose $T=3\cdot 10^{-5}$ s, $P=3$, $h=5\cdot 10^{-4}$ m and 2000 time steps.\\
Figure \ref{fig:absorbing} depicts four snapshots of the simulation and qualitatively shows the well-behaviour of the absorbing imposition. \\
In order to quantify the perturbation due to the ultrasounds in the domain, we compute and display in Figure \ref{fig:decay} the $L^2$ norm of the sound pressure at each time step. We observe that the perturbation starts from zero and reaches a constant value, that is kept from roughly $t=0.5\cdot 10^{-5}$ s to $t=1.25\cdot 10^{-5}$ s. In this time interval there is perfect conservation of the energy in the domain since there is no further source term and the absorbing walls have not been reached yet. The norm of the pressure starts to decrease when the waves hit the absorbing walls at around $t=1.25\cdot 10^{-5}$ s. The steepest descent coincides with the absorption of the main horizontal component, while the pressure perturbation decreases more gradually when the waves are later absorbed in the other directions. In the final part we observe the quality of the proposed algorithm: while the orthogonality assumption (i.e., Equation (\ref{absorbing wall}) with $\theta=0$) effectively removes the main wave components, it is the correct estimate of the incident angles from Equation (\ref{filtering}) that allows to remove the remaining energy from the domain.

\subsection{A 2D clamp-on flow meter simulation}\label{A 2D clamp-on flow meter}
This test evaluates the proposed methods in a more practical context such as the measurement in a 2D clamp-on flow meter. While this setup is far from comparable to modern industrial flow meters, which have a high level of three-dimensional complexity, it represents a fair compromise to analyse the behaviour of the methods in a simple setting. Moreover, similar geometries have been previously adopted in other works such as \cite{luca2016numerical}.

\begin{figure}[ht]
\centering
  \begin{subfigure}[t]{.40\textwidth}
  \centering
    \includegraphics[width=\linewidth]{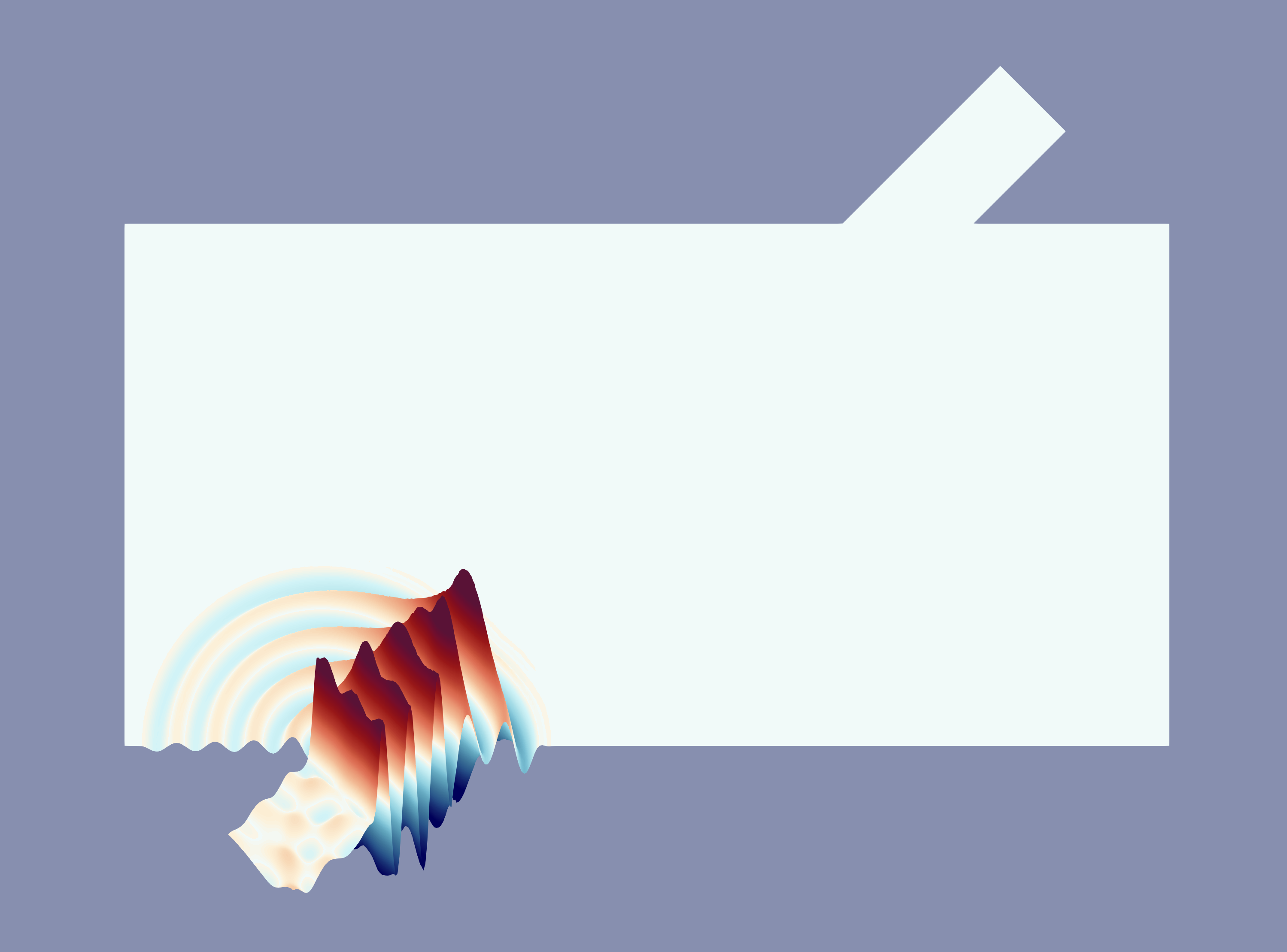}
    \makebox[1cm][l]{%
    \raisebox{0.5cm}[0pt][0pt]{\hspace{2cm}
    \includegraphics[width=0.08\linewidth]{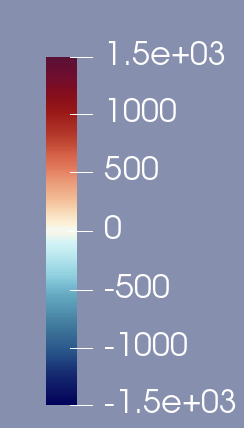}}}
    \caption{$t=8\cdot10^{-6}$ s}
  \end{subfigure}
  \hspace{0.06\textwidth}
  \begin{subfigure}[t]{.40\textwidth}
  \centering
    \includegraphics[width=\linewidth]{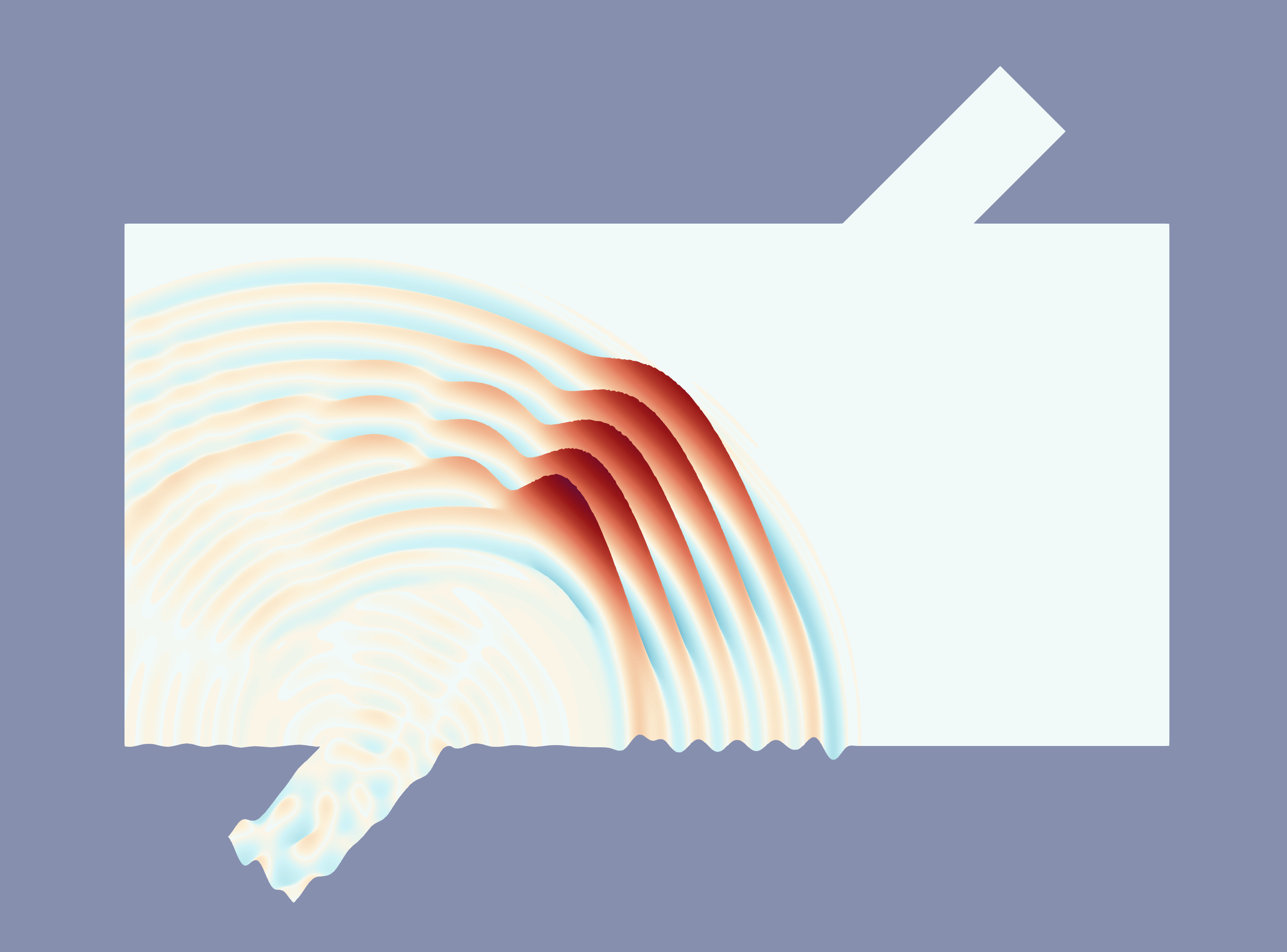}
    \caption{$t=1.6\cdot10^{-5}$ s}
  \end{subfigure}
  \medskip

  \begin{subfigure}[t]{.40\textwidth}
    \centering
    \includegraphics[width=\linewidth]{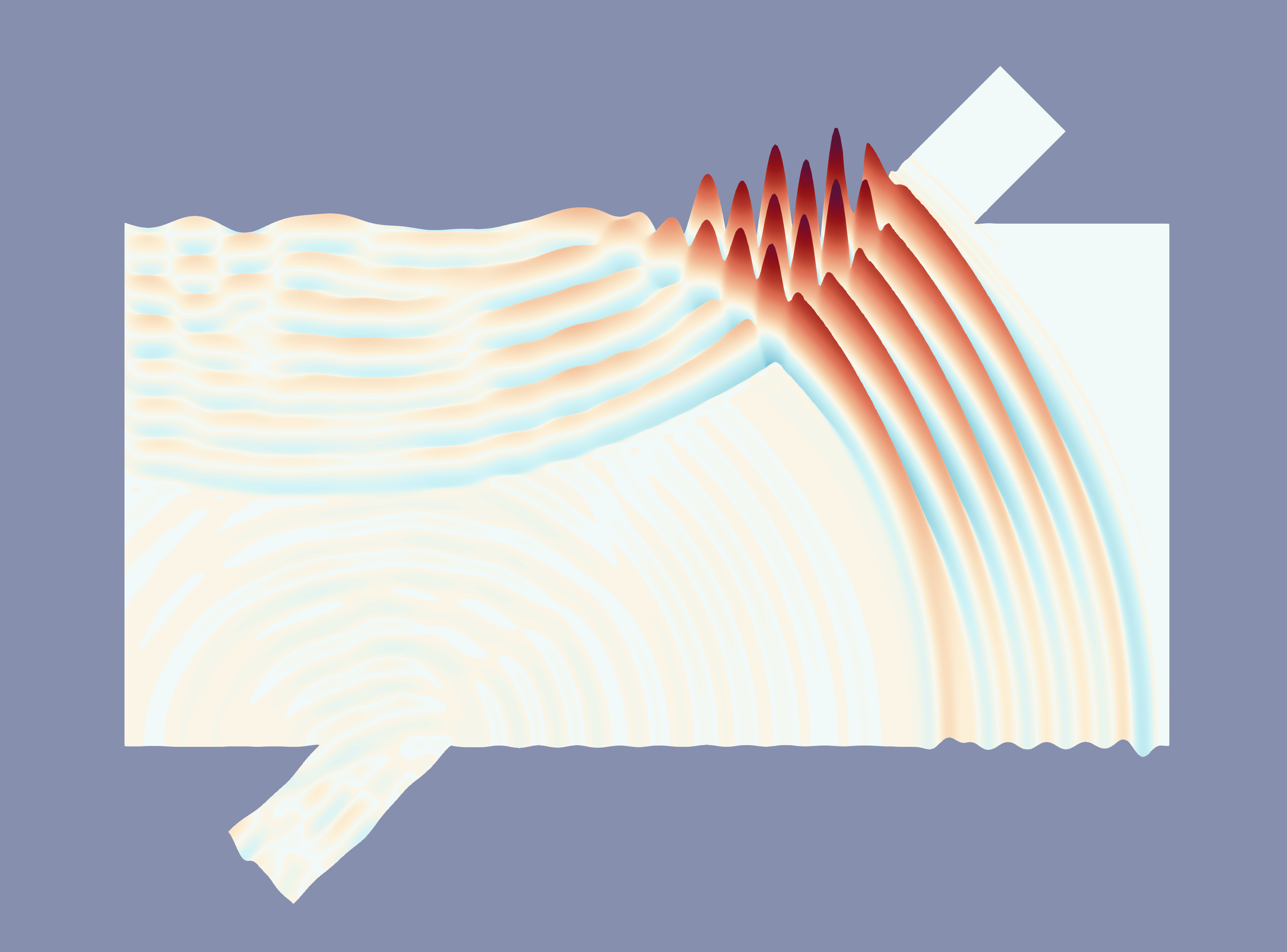}
    \caption{$t=2.4\cdot10^{-5}$ s}
  \end{subfigure}
  \hspace{0.06\textwidth}
  \begin{subfigure}[t]{.40\textwidth}
    \centering
    \includegraphics[width=\linewidth]{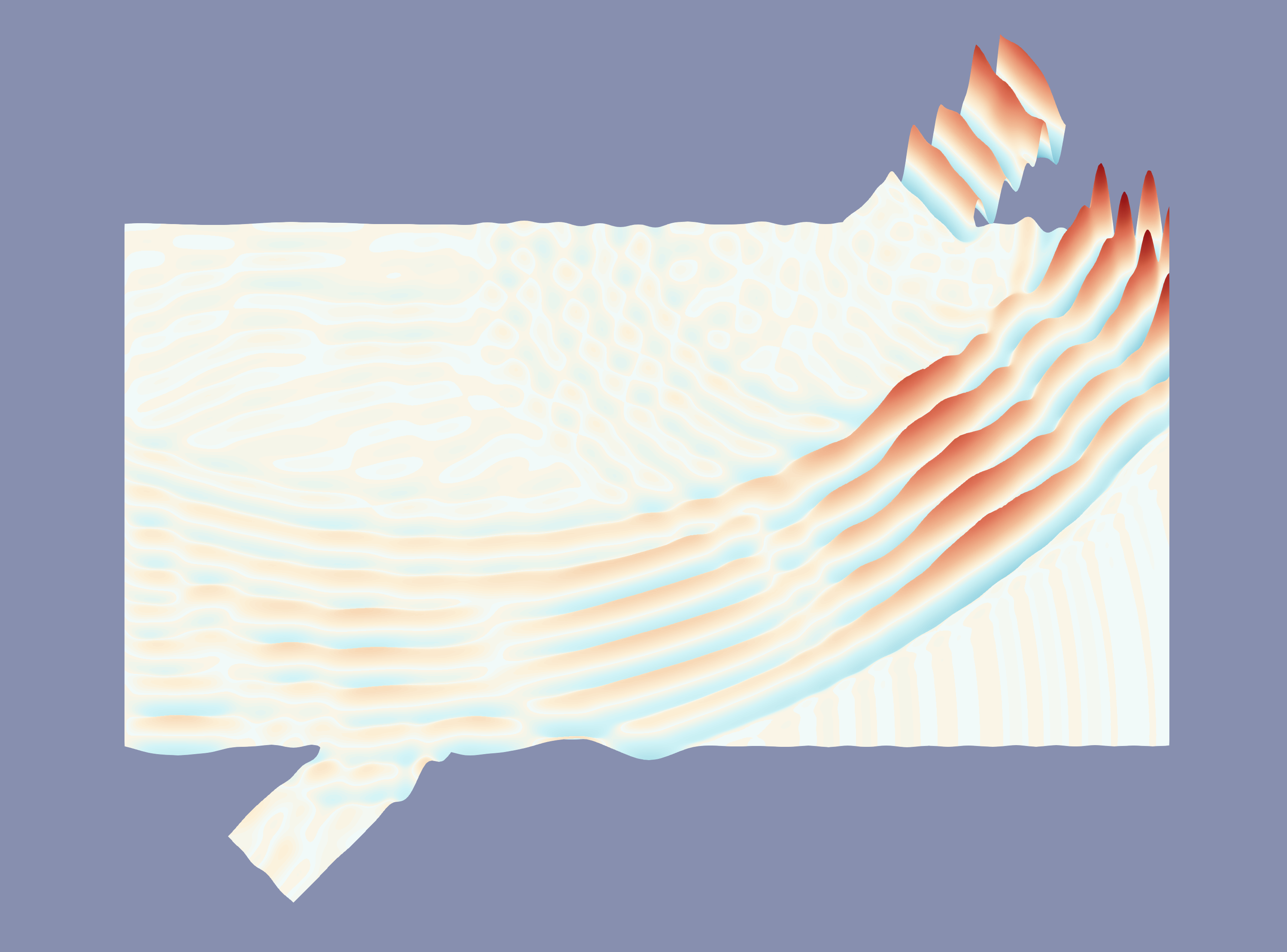}
    \caption{$t=3.2\cdot10^{-5}$ s}
  \end{subfigure}
  \caption{Perturbation pressure snapshots from the clamp-on test proposed in Section \ref{A 2D clamp-on flow meter}. The pressure values range in $\pm1.5\cdot10^3$ Pa.}
  \label{clampon}
\end{figure}

\begin{figure}[ht]
\centering
  \begin{subfigure}[t]{0.8\textwidth}
    \centering
    \includegraphics[width=\linewidth]{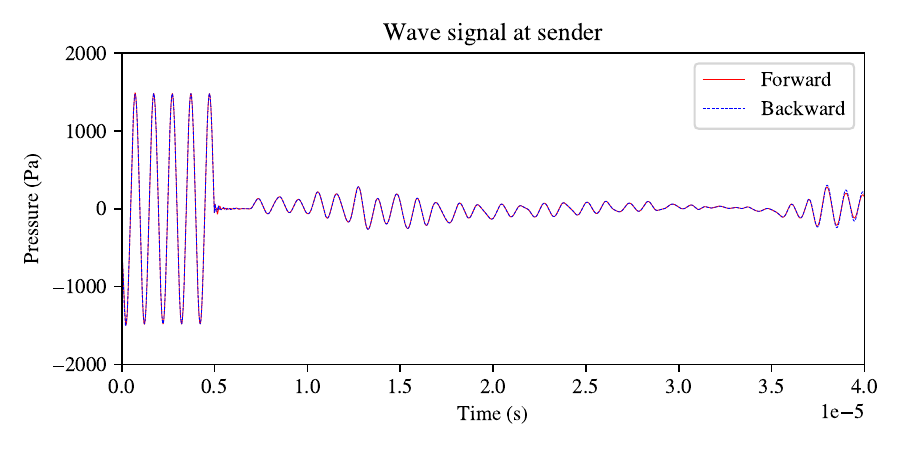}
  \end{subfigure}
  \medskip
  \begin{subfigure}[t]{0.8\textwidth}
    \centering
    \includegraphics[width=\linewidth]{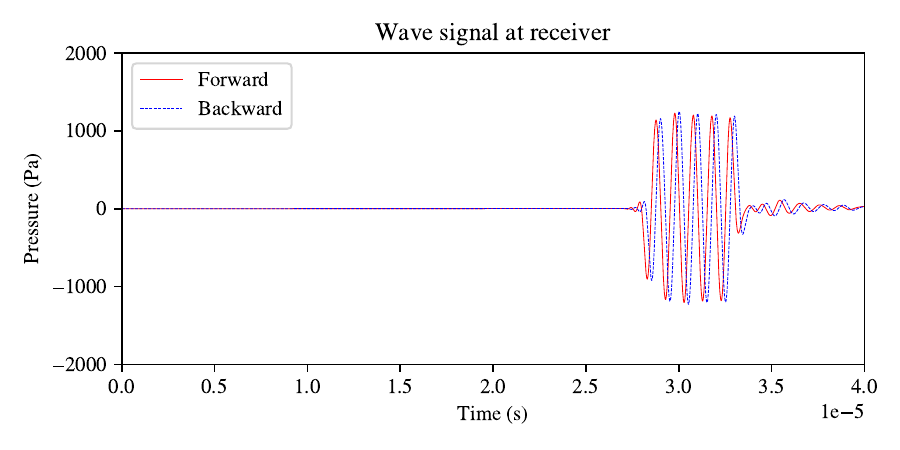}
  \end{subfigure}
  \caption{Pressure measurements at the two transducers for the forward and backward waves. As expected, the two waves overlap at the sender but they are not completely aligned at the receiver because of the water convection.}
  \label{pressure_measurement}
\end{figure}

The simple clamp-on geometry is shown in Figure \ref{clampon} where the horizontal and vertical lengths are respectively 0.04m and 0.02m. We choose $T=4\cdot10^{-5}$ s, $P=3$, $h=4\cdot 10^{-4}$ m and 3000 time steps. The same inlet profile from Equation (\ref{test_inlet}) is applied at the bottom-left corner. Moreover, the same Poiseuille background flow from Equation (\ref{poiseuille}) is used. We assign perfectly-reflecting walls on the entire boundary except for the absorbing layers on the vertical walls (that represent the flow inlet and outlet) with $\alpha_M=0.01$ and the top-right corner which is imposed as a realistic/resistive wall with perpendicular incident angle. In particular, assuming the flow meter to be made of polyphenylene sulfide (PPS), we assign $\rho_w=1650$ kg/m$^3$ and $c_w=2800 $m/s. The problem is hence solved non-dimensionally with respect to the SI system.\\ 
Qualitative snapshots are shown in Figure \ref{clampon}. The behaviour of the walls and absorbing layers are as expected. While not shown, the plots of the velocity and density perturbation have similar characteristics. \\
Besides these qualitative results, it is interesting to investigate the pressure measurements at the transducers. Trying to mimic the functioning of an ultrasonic transit-time flow meter, a second simulation is performed where the acoustic wave inlet and the resistive wall are switched. Therefore, we distinguish between forward (moving with the flow) and backward (moving against the flow) waves as well as bottom-left and top-right transducers. The measurements of both waves are then shown in Figure \ref{pressure_measurement}. 
While at the sender the two pressure measurements are identical, the two waves presents a phase shift when they are measured on the receiving transducer because of the wave convection due to the water flow. The temporal difference between the wave peaks of the two signals is on average 18 time steps, i.e., $2.4\cdot10^{-7}$ s. Hence, we can use the classical formula for the estimation of the water velocity in transit-time flow-meters \cite{FlowMeasurementHandbook}
\begin{equation*}
    \hat{v} = \frac{c^2 \Delta t}{2 \Delta x},
\end{equation*}
where $\Delta t$ is the difference in time between forward and backward waves and $\Delta x$ is the horizontal distance between sender and receiver. The obtained result is $\hat{v}\approx 8.9$ m/s that is indeed a precise estimate of the averaged horizontal background velocity $\frac{1}{L}\int_\gamma \bar{u}(x,y) \approx 9.0178$ m/s, where $\gamma$ is the path from the sender to the receiver and $L$ its length.

\subsection{Parallel calculations for the 3D flow meter simulation} \label{Parallel calculations}
\begin{figure}[ht]
\centering
  \begin{subfigure}[t]{.36\textwidth}
  \centering
    \includegraphics[trim={0 6cm 0 6cm}, clip, width=\linewidth]{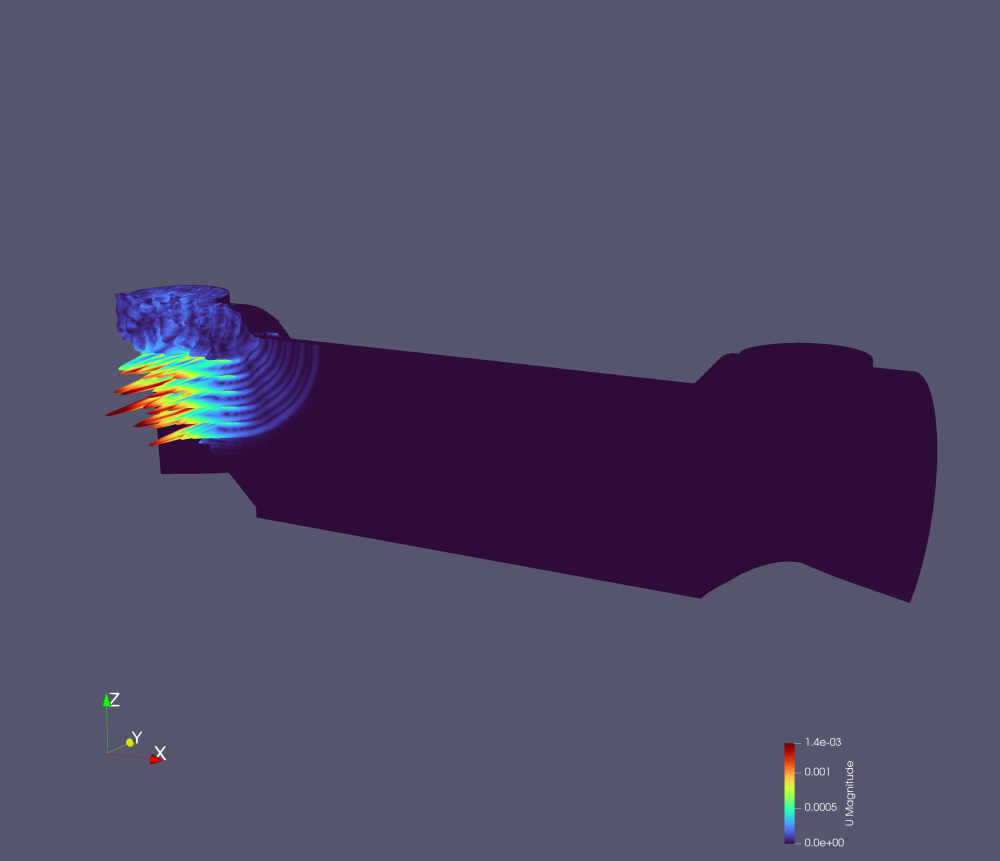}

\makebox[1cm][l]{%
  \raisebox{0.5cm}[0pt][0pt]{\hspace{-1.7cm}
    \includegraphics[width=0.08\linewidth]{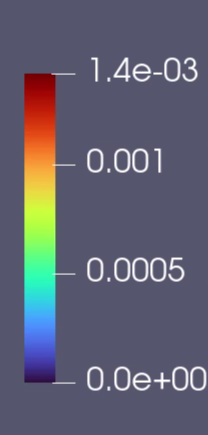}}}
          
    \caption{$t=8\cdot10^{-6}s$}
  \end{subfigure}
  \hspace{0.06\textwidth}
  \begin{subfigure}[t]{.36\textwidth}
  \centering
    \includegraphics[trim={0 6cm 0 6cm}, clip, width=\linewidth]{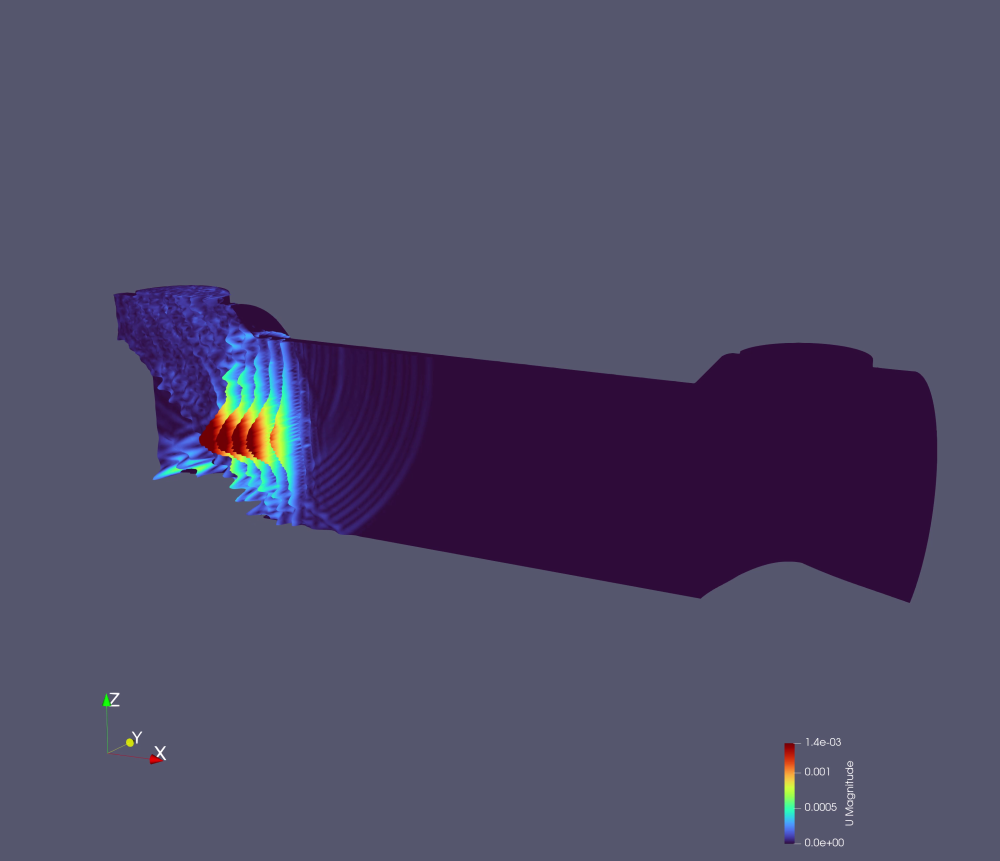}
    \caption{$t=1.6\cdot10^{-5}s$}
  \end{subfigure}
  \medskip

  \begin{subfigure}[t]{.36\textwidth}
    \centering
    \includegraphics[trim={0 6cm 0 6cm}, clip, width=\linewidth]{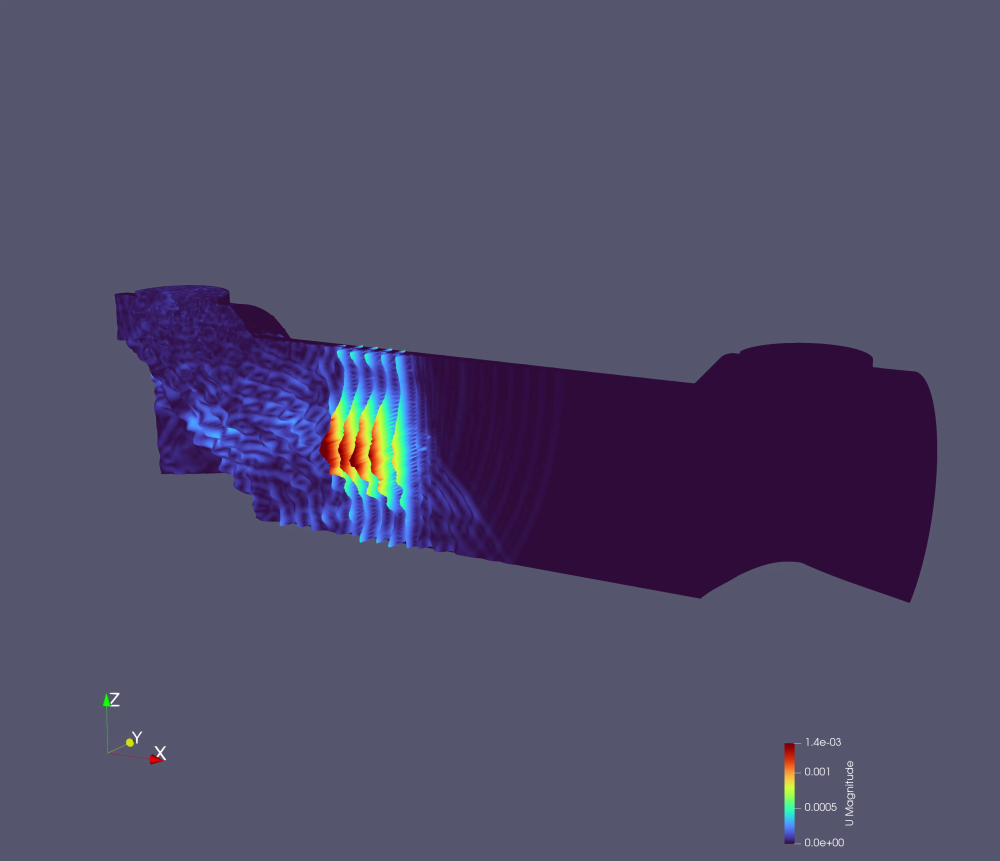}
    \caption{$t=2.4\cdot10^{-5}s$}
  \end{subfigure}
  \hspace{0.06\textwidth}
  \begin{subfigure}[t]{.36\textwidth}
    \centering
    \includegraphics[trim={0 6cm 0 6cm}, clip, width=\linewidth]{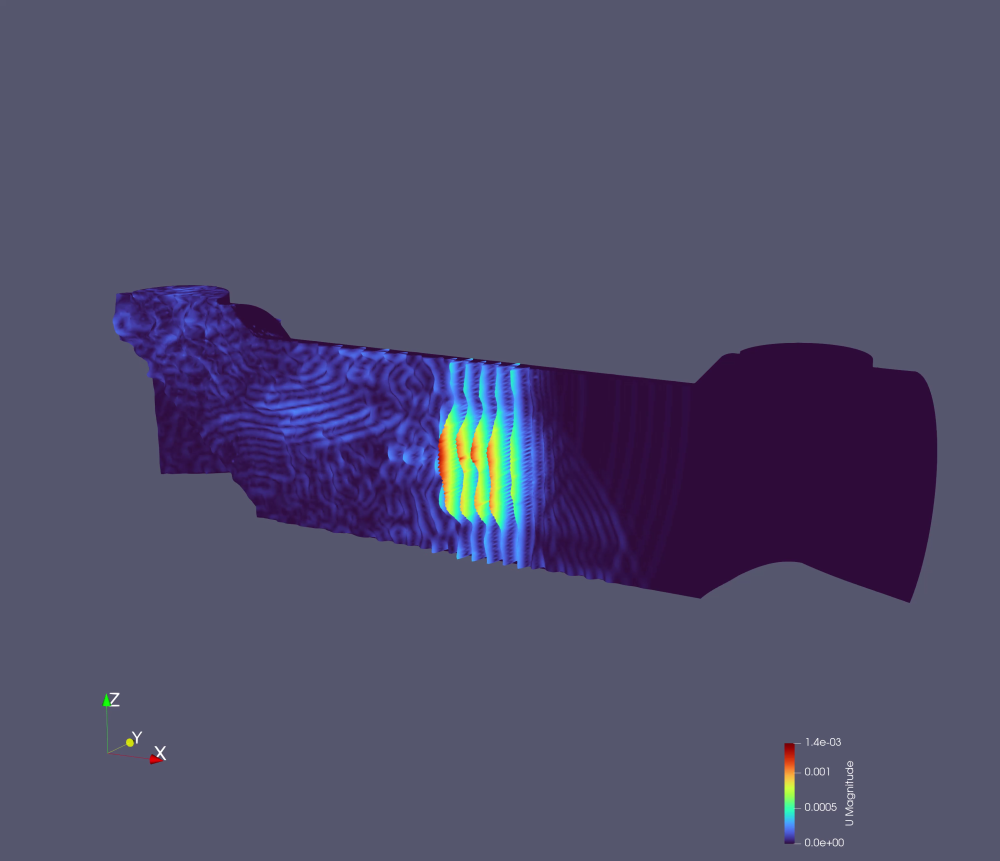}
    \caption{$t=3.2\cdot10^{-5}s$}
  \end{subfigure}

  \medskip

  \begin{subfigure}[t]{.36\textwidth}
    \centering
    \includegraphics[trim={0 6cm 0 6cm}, clip, width=\linewidth]{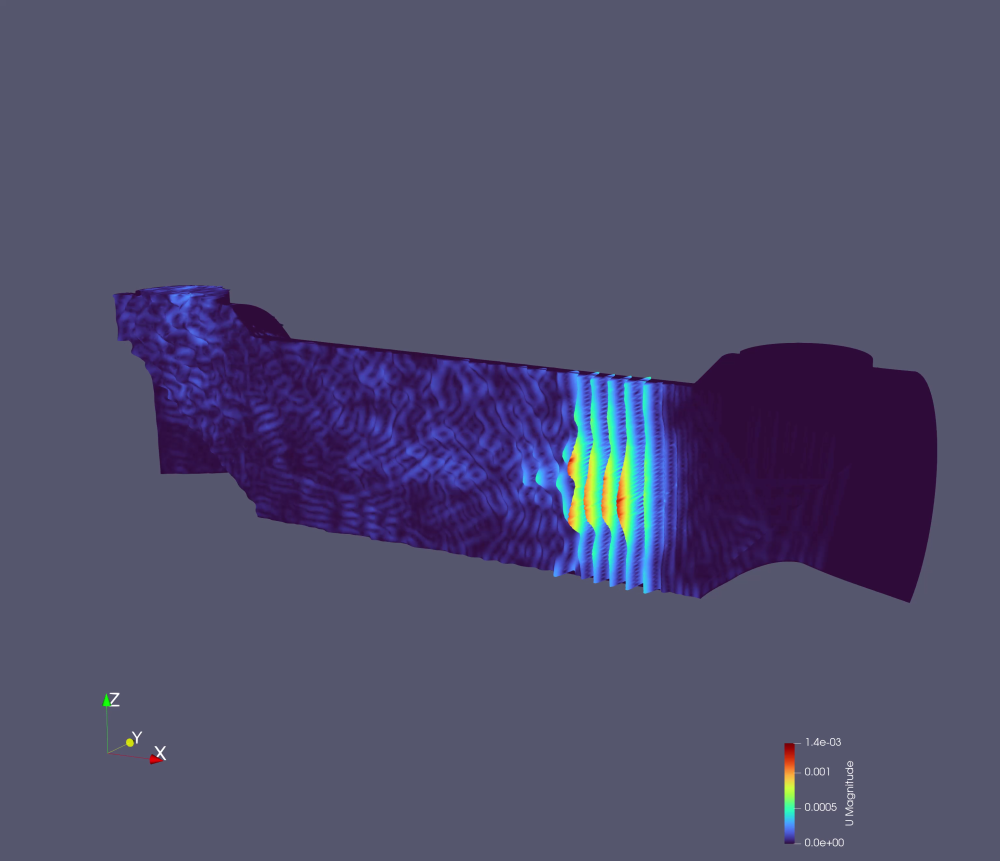}
    \caption{$t=4.0\cdot10^{-5}s$}
  \end{subfigure}
  \hspace{0.06\textwidth}
  \begin{subfigure}[t]{.36\textwidth}
    \centering
    \includegraphics[trim={0 6cm 0 6cm}, clip, width=\linewidth]{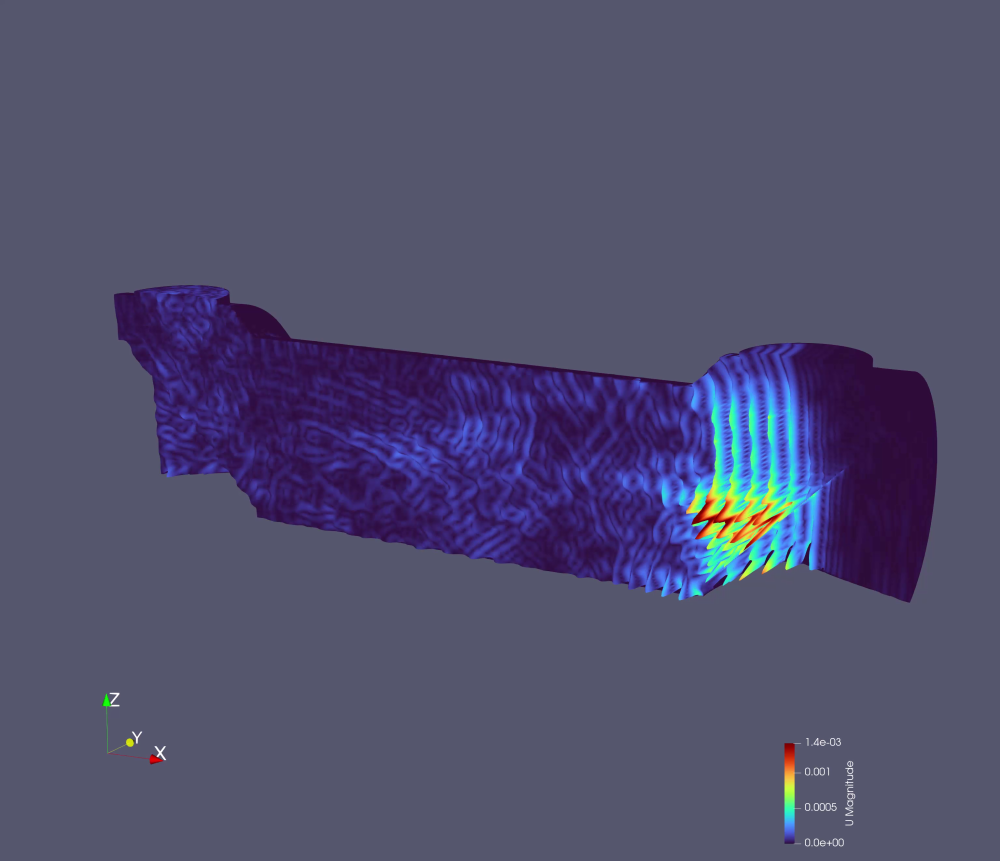}
    \caption{$t=4.8\cdot10^{-5}s$}
  \end{subfigure}

  \medskip

  \begin{subfigure}[t]{.36\textwidth}
    \centering
    \includegraphics[trim={0 6cm 0 6cm}, clip, width=\linewidth]{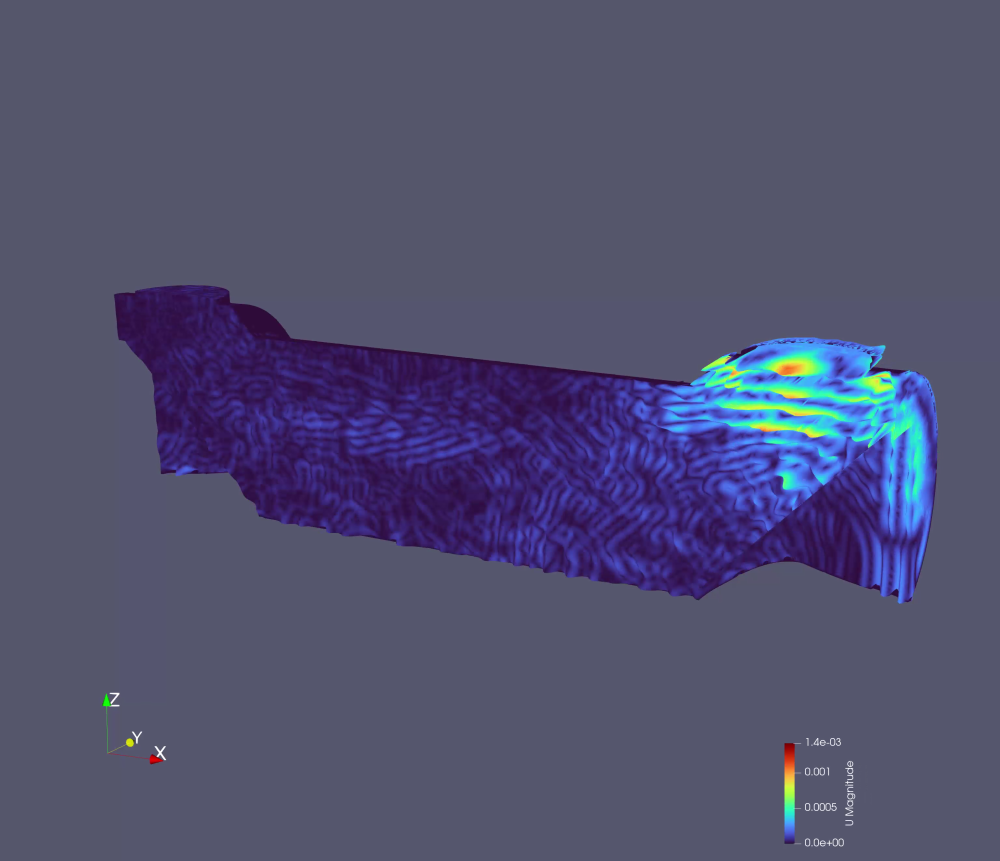}
    \caption{$t=5.6\cdot10^{-5}s$}
  \end{subfigure}
  \hspace{0.06\textwidth}
  \begin{subfigure}[t]{.36\textwidth}
    \centering
    \includegraphics[trim={0 6cm 0 6cm}, clip, width=\linewidth]{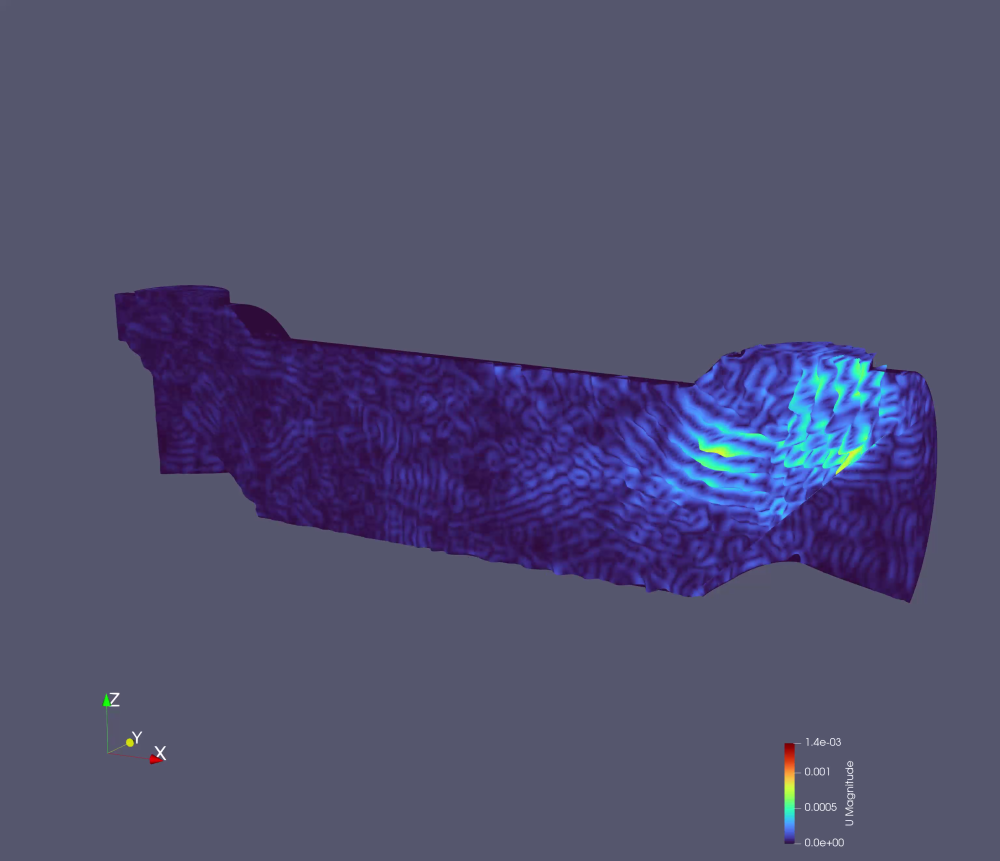}
    \caption{$t=6.4\cdot10^{-5}s$}
  \end{subfigure}
  \caption{Some screenshots from the simulation discussed in Section \ref{Parallel calculations}. Colors refer to different values of the velocity magnitude $U=\sqrt{\mathbf{u}_1^2+\mathbf{u}_2^2+\mathbf{u}_3^2}$ which range from $0$ m/s to $1.4\cdot10^{-3}$ m/s.}
  \label{3d_sim}
\end{figure}

Previous tests and simulations are performed only in two spatial dimensions and simple geometries. Therefore, this final section is dedicated to a qualitative assessment of the correct 3D extension of the previously proposed methods as well as the application to a more realistic case. Calculations are in this latter case more demanding and so we introduce a simple but powerful parallel scheme.\\
The most expensive operations at each time step are represented by the $\mathcal{S}$ matrix assembly and the multiple matrix-vector products using $\mathcal{S}$. Indeed, no system resolution is performed thanks to the pre-inversion of the mass matrix and the cost to assembly the remaining terms (vectors and the diagonal mass matrix) is in comparison negligible. A very effective parallelization consists in the assembly of local $\mathcal{S}_r$ sparse matrices, where $r=1,\dots,R$ is the rank number, each one referring to a subset of the entire domain yielding a domain decomposition method. To have more efficient calculations and no further memory occupied, it is preferred to have a non-overlapping partition where the shared interfaces have minimum length. The domain splitting algorithm has been implemented over FreeFem++ and is based on selecting vertical slices with regular horizontal length. This simple strategy is motivated by the horizontally extended geometry of the ultrasonic meters but there is certainly room for improvement. The local matrices are never assembled together, instead the matrix-vector product is performed with the local matrix and the global vector. Finally, an \texttt{MPI\_reduce} operation allows to sum up all the local vector results in order to obtain the global product. This simple algorithm turns out to be very powerful since it almost-linearly reduces the costs related to the two mentioned expensive operations. \\
Adopting this strategy with $R=20$ cores, we perform a 3D simulation taking the geometry from \cite{rincon2022turbulent}, which is described as a suitable compromise between numerical applications and fidelity to the \emph{lowIQ 2200} meter from Kamstrup A/S. Different from the simpler clamp-on geometry in Section \ref{A 2D clamp-on flow meter}, the sender and receiver are placed on the upper border of the pipe while the correct trajectory is given from the double reflection on two $45^{\circ}$ wedges located on the bottom. To reduce the calculation effort, only one of the two halves of the pipe is used imposing a symmetry condition $\Phi=0$. Moreover, the two lateral faces are again imposed as absorbing with $\alpha_M=0.01$ and the receiver is imposed as a PPS resistive wall with perpendicular incident angle. The wedges are instead imposed as perfectly reflecting since the $45^\circ$ angle is greater than the critical angle. The source profile is the same as in Equation (\ref{test_inlet}), the background flow is given from the RANS $k-\epsilon$ solution in \cite{rincon2022turbulent} and $T=8\cdot 10^{-5}$ s with $M=10^4$ time steps. To obtain a high fidelity simulation, the mesh is composed by 248'646 tetrahedral elements, which correspond to the averaged grid size $h=0.55\cdot10^{-3}$ m, and the polynomial order is again $P=4$. We notice that these choices lead to the significantly large value $N_{h,P} \approx 5\cdot 10^6$ for each of the four variables. This can be considered as highly demanding for standard FEM schemes but the parallel RKDG method takes approximately 28 hours and the results are then shown in Figure \ref{3d_sim}. Thus, we can qualitatively confirm the capacity of the method to correctly simulate the entire trajectory of the ultrasound wave followed by its dissipation in a reasonable time frame. 

\section{Conclusions}
In this work we have simulated the propagation of ultrasonic waves through the approximation of the linearized Euler equations by discontinuous Galerkin methods with the Dubiner spectral basis and a novel algorithm for the imposition of absorbing and resistive walls. \\
Several tests performed to assess this approach showed good convergence behaviour and the expected performance improvement due to the use of the spectral basis and a Runge-Kutta explicit-time formulation. Moreover, when combined with the proposed absorbing conditions, this method proves itself capable of accurately and efficiently simulating also more complex conditions like the sound wave propagation within an ultrasonic flow meter. \\
Hence, despite the simplicity of the proposed tests, we believe that such studies can be easily extended to more practical problems and that the industrial research community can exploit these advances in numerical methods for conservation laws to improve the precision of the modern flow metering instruments.

\section{Data availability}
Data and code are properties of Kamstrup A/S. They are available upon request under certain conditions.

\section*{Acknowledgments}
The authors aim to thank professor Jan Hesthaven (EPFL) for his helpful guidance on the theory of DG methods. An important contribution came also from the frequent discussions with Lasse Nielsen and Mario Rincón, collaborators at Kamstrup A/S.

\appendix
 \section{Dubiner-Koornwinder polynomials} \label{Dubiner-Koornwinder polynomials}
The Dubiner-Koornwinder basis has been firstly presented by Koornwinder \cite{koornwinder1975two}, re-introduced by Dubiner \cite{Dubiner} and extended in 3 dimensions by Sherwin and Karniadakis \cite{sherwin}. We dedicate this section to briefly define such functions using the same notation from \cite{antonietti-houston}.
 \begin{definition}[Jacobi polynomials \cite{jacobi}]\label{jacobi}
    The Jacobi polynomials $\{P_n^{\alpha,\beta}\}_{n\in \mathbb{N}_0}$ of indices $\alpha,\beta \in \mathbb{R}$ are the class of polynomials that are orthogonal in $L^2(1,1)$ with respect to the Jacobi weight $w_{\alpha,\beta}(x)=(1-x)^\alpha(1+x)^\beta$:
    \begin{equation*}
    \int_{-1}^{1}w_{\alpha,\beta}(x) P_n^{\alpha,\beta}(x) P_m^{\alpha,\beta}(x) \, dx=\frac{2^{\alpha+\beta+1}\Gamma(n+\alpha+1)\Gamma(n+\beta+1)}{(2n+\alpha+\beta+1)\Gamma(n+\alpha+\beta+1)n!}\delta_{nm},
    \end{equation*}
    where $\delta_{nm}$ is the Kronecker delta function.
    They can be explicitly computed as
    \begin{equation*}
        P_n^{\alpha,\beta}(x)= \frac{(-1)^n}{2^n n!} (1-x)^{-\alpha}(1+x)^{-\beta} \frac{d^n}{dx^n} \left[(1-x)^\alpha(1+z)^\beta(1-z^2)^n\right].
    \end{equation*}
    \end{definition}
    
    \begin{definition}[2D Dubiner basis \cite{antonietti-houston}]
    If $\hat{K} = \{(\xi,\eta): \xi,\eta\ge 0, \xi+\eta \le 1\}$ is the reference triangle, the 2D Dubiner basis function indexed by $(i,j) \in \mathbb{N}_0^2$ is defined as:
    \begin{gather*}
        \varphi_{i,j}:\hat{K}\rightarrow\mathbb{R}\\
        \varphi_{i,j}(\xi,\eta)=c_{i,j} (1-\eta)^i P_i^{0,0}\left(\frac{2\xi}{1-\eta} -1\right) P_j^{2i+1,0}(2\eta-1),
    \end{gather*}
    where $c_{i,j}=\sqrt{2(2i+1)(i+j+1)}$.
    \end{definition}
    \begin{definition}[3D Dubiner basis \cite{sherwin}]\label{3D Dubiner basis}
    If $\hat{V} = \{(\xi,\eta,\nu): \xi,\eta,\nu\ge 0, \xi+\eta+\nu \le 1\}$ is the reference tetrahedron, the 3D Dubiner basis function indexed by $(i,j,k) \in \mathbb{N}_0^3$ is defined as:
    \begin{equation*}
    \varphi_{i,j,k}:\hat{V}\rightarrow\mathbb{R}
    \end{equation*}
    \begin{equation*}
    \begin{aligned}
        \varphi_{i,j,k}(\xi,\eta,\nu)= &c_{i,j,k} P_i^{0,0}\left(\frac{2\xi}{1-\eta-\nu}-1\right) \\ &\left(1-\eta-\nu \right)^i P_j^{2i+1,0}\left(\frac{2\eta}{1-\nu} -1\right) \\ &\left(1-\nu\right)^j P_k^{2i+2j+2,0}(2\nu-1),
    \end{aligned}
    \end{equation*}
    where $c_{i,j,k}=\sqrt{(2i + 2j + 2k + 3)(2i + 2j + 2)(2i + 1)}$.
    \end{definition}

        \begin{proposition}[Orthonormality of the Dubiner basis] $\{\varphi_{i,j}\}_{i+j\le P}$ and $\{\varphi_{i,j,k}\}_{i+j+k\le P}$ are $L^2$-orthonormal bases of $\mathbb{P}^P(\hat{K})$ and $\mathbb{P}^P(\hat{V})$, respectively, i.e., the space of polynomials of degree at most $P\in\mathbb{N}_0$.
    \end{proposition}
    
    \begin{proof}
    The proofs are available in \cite{sherwin} (Section 3.1 and 3.2) but need to be adapted to a different reference triangle and tetrahedron. 
    \end{proof}

\bibliographystyle{abbrv}
\bibliography{bibliography}

\end{document}